%% file: main.tex
\documentclass[conference]{IEEEtran}

\ifCLASSINFOpdf
\else
\fi

\usepackage{tikz}
\usepackage{amsmath}

\usepackage{filecontents}

\usepackage{cite}

\usepackage{textcomp}

\usepackage{xurl}

\usepackage{url}
\usepackage{cite}
\usepackage{tikz}
\usepackage{amsmath}

\usepackage{xcolor}
\usepackage{microtype}
\microtypecontext{spacing=nonfrench}
\usepackage[colorlinks=true,
            linkcolor=blue,
            citecolor=blue,
            urlcolor=blue]{hyperref}
\usepackage{cleveref}
\usepackage[most]{tcolorbox}
\usepackage{enumitem}
\usepackage{xspace}
\usepackage{subfigure}
\usepackage{cleveref}
\usepackage{booktabs}
\usepackage{threeparttable} 
\usepackage{graphicx}
\usepackage{pifont}
\usepackage{caption}
\usepackage{mdframed} 
\usepackage{tcolorbox}
\usepackage{float} 
\usepackage{makecell}
\usepackage{multirow}
\usepackage{array}
\usepackage{color}
\usepackage{booktabs,tabularx}
\crefname{figure}{Fig.}{Figs.}
\crefname{table}{Table}{Tables}
\crefname{section}{Sec.}{Secs.}
\crefname{equation}{Eq.}{Eqs.}
\crefname{appendix}{App.}{Apps.}
\newcommand{\rev}[1]{\textcolor{black}{#1}}
\newenvironment{revision}{\color{black}}{}

\newcommand{\xxx}{SenTrojan\xspace}
\usepackage{booktabs} 
\usepackage{arydshln} 
\begin{document}
%
 \title{\xxx: Revealing Analog Hardware Trojan Attacks on Sensors}
\author{%
    Kun~Wang\textsuperscript{1},
    Ruochen~Zhou\textsuperscript{2,*},
    Wenjun~Zhu\textsuperscript{1},
    Shilin~Xiao\textsuperscript{1},\\
    Kai~Wang\textsuperscript{1},
    Chen~Yan\textsuperscript{1},
    Xiaoyu~Ji\textsuperscript{1},
    and~Wenyuan~Xu\textsuperscript{1}\\[0.3em]
    {\normalsize
        \textsuperscript{1}Zhejiang University, China,
        \{eewk, zwj\_, xshilin, eekaiwang, yanchen, xji, wyxu\}@zju.edu.cn
    }\\
    {\normalsize
        \textsuperscript{2}The Hong Kong University of Science and Technology, China, zrccc@ust.hk
    }%
    \thanks{\textsuperscript{*}~Ruochen Zhou is the corresponding author.}%
}

\IEEEoverridecommandlockouts
\makeatletter
\def\@IEEEpubidpullup{6.5\baselineskip}
\makeatother


\maketitle

\input{sections/abstract}

\input{sections/introduction}

\input{sections/background}

\input{sections/threatmodel}

\input{sections/preliminary_analysis}
\input{sections/design}

\input{sections/Evaluation}

\input{sections/discussion}
\input{sections/related_work}

\input{sections/conclusion}

\bibliographystyle{unsrt}
\bibliography{reference/reference.bib}

\appendix

\input{sections/Appendix}

\end{document}

%% file: sections/abstract.tex
\begin{abstract}

Sensors bridge the physical and digital worlds, making their integrity critical to cyber-physical system security. 
Existing sensor attacks largely exploit unintended and sensor-specific physical vulnerabilities, limiting their generalizability and motivating a different threat model, i.e., whether an adversary can implant hidden attack surfaces into sensors before deployment. 
In this paper, we present \xxx, an analog hardware Trojan framework for sensors. 
\xxx embeds malicious functionality into analog signal circuitry commonly shared across diverse sensor architectures, and activates the hidden behavior through crafted external physical signals. 
Once activated, \xxx can induce sensor malfunctions ranging from permanent denial-of-service to static or dynamic spoofing. 
To achieve these effects under different architectural constraints, we introduce two complementary designs. 
The Logic-Rule-Driven (LRD) Trojan adds lightweight analog monitoring logic to detect trigger features such as frequency and amplitude, enabling reliable activation across various sensor types. 
The Physical-Effect-Driven (PED) Trojan targets sensors with exploitable native analog effects, e.g., nonlinearity and rectification, and realizes attacks with minimal component insertion. 
We evaluate \xxx on 14 sensors spanning 6 transduction principles and demonstrate its effectiveness across diverse real-world scenarios. 
Overall, \xxx achieves an average successful trigger rate above 95\%, maintains over 97\% similarity to benign behavior, incurs zero false-trigger rate for LRD designs, and introduces less than 5\% average deviation in drift, noise floor, and response time. 
To our knowledge, this work is the first systematic study of analog sensor Trojans, broadening sensor security from unintended physical vulnerabilities to intentional supply-chain compromise.

\end{abstract}

%% file: sections/introduction.tex
\section{Introduction}

Sensors serve as the critical interface between physical environments and digital systems, underpinning a wide range of applications, including autonomous vehicles, industrial automation, and healthcare monitoring. The reliability and safety of these systems depend fundamentally on correct sensor operation. Catastrophic consequences resulting from sensor failures, such as airplane crashes~\cite{curt2019boeing}, automotive accidents~\cite{stewart2018tesla}, and industrial disasters~\cite{process2018excess}, underscore the importance of sensor security.

Existing sensor security research has predominantly focused on exploiting inherent vulnerabilities through signal injection attacks. These approaches leverage unintentional hardware imperfections, e.g., circuit nonlinearity~\cite{zhang2017dolphinattack} and electromagnetic (EM) susceptibility~\cite{kune2013ghost}, to compromise sensor readings. 
However, because these vulnerabilities arise unintentionally during design or implementation, they are often inconsistent across sensor models, hardware versions, and deployment environments. 
As a result, existing attacks typically require pre-existing exploitable weaknesses, and substantial effort has been devoted to characterizing and detecting them~\cite{phyfuzz}.

\begin{figure}[t]
    \centering
    \includegraphics[width=1.0\linewidth]{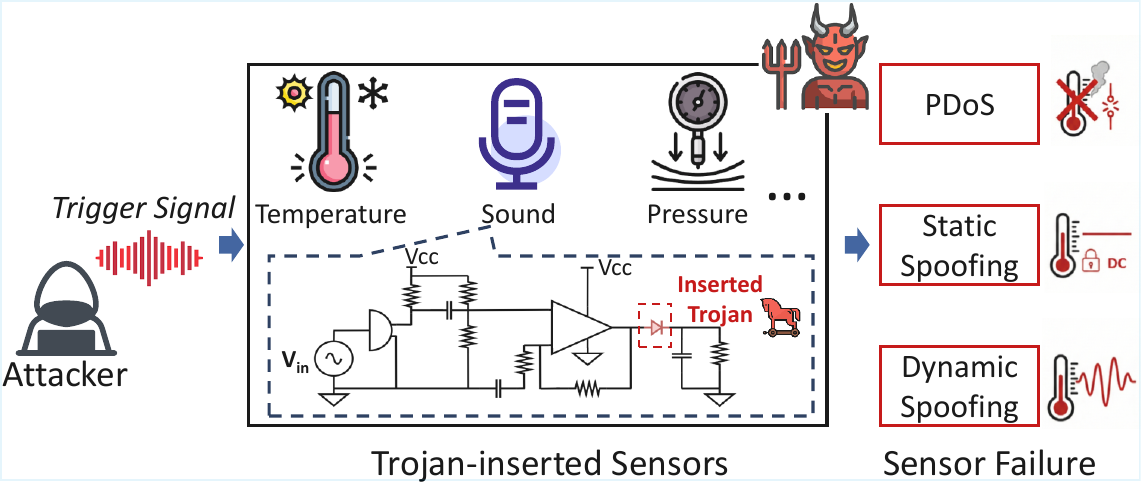}
    \caption{An illustration of \xxx. The attacker can use a pre-designed trigger signal to activate the analog sensor Trojan, leading to PDoS, static spoofing, or dynamic spoofing.}
    \label{fig1}
\end{figure}
Rather than passively searching for existing sensor vulnerabilities, we investigate an alternative threat model, i.e., the deliberate insertion of exploitable vulnerabilities into sensors before deployment. 
This threat is supported by real-world precedents and established digital paradigms, e.g., hardware Trojans~\cite{bhunia2014hardware,xiao2016hardware,puschner2023red,dharsee2023jinn} and neural network backdoors~\cite{li2022backdoor}.
However, this paradigm remains largely unexplored in analog sensor circuitry. 
Although prior work has explored Trojan designs using analog components~\cite{yang2016a2}, these designs primarily replace digital counters with analog capacitors, while still monitoring digital states and relying on specific data patterns or instructions for activation. 
Their triggers are therefore ultimately software-issued binary events rather than physical signals received through the sensing interface.
This leaves open the central challenges of analog sensor Trojans, i.e., \textit{how to preserve the accuracy of the analog front end while achieving reliable Trojan activation under noisy physical inputs}.


In this paper, we introduce \xxx, the first supply-chain analog hardware Trojan framework for sensors. Unlike prior sensor attacks that passively exploit pre-existing physical vulnerabilities, we examine an active compromise model in which an adversary deliberately inserts exploitable analog behavior into the sensor circuitry before deployment. 
The attacker can activate the hidden Trojan using a crafted external physical signal, without requiring software privileges or digital access to the victim system, as shown in~\cref{fig1}.
\xxx expands the sensor attack surface in two ways. First, it enables attack effects that are difficult to achieve through direct signal injection alone, such as persistent output manipulation.
Second, it can compromise sensors that are resistant to known signal injection attacks or amplify weak physical vulnerabilities that would otherwise have limited attack impact.



To support these attack scenarios under different sensor architectures and insertion constraints, \xxx introduces two complementary designs. 
The Logic-Rule-Driven (LRD) Trojan adds lightweight analog circuitry to monitor crafted physical signals and control malicious payloads. By checking features such as frequency, amplitude, and duration against predefined rules, LRD Trojans provide reliable activation across diverse sensor types and support flexible payload behaviors.
The Physical-Effect-Driven (PED) Trojan further targets sensors whose native analog effects can be induced or amplified by crafted inputs.
PED Trojans avoid explicit monitoring logic and instead tune existing analog components so that the sensor’s own physical behavior serves as the trigger and payload mechanism. This design enables minimal hardware changes while inducing exploitable vulnerabilities in resistant sensors or strengthening weak existing vulnerabilities.

However, realizing \xxx in the analog sensor domain requires addressing a common challenge of functionality and stealth, i.e., a Trojan must reliably respond to attacker-controlled physical triggers while preserving normal sensor behavior before activation. 
For LRD Trojans, the main challenge is reliable and selective triggering without perturbing the sensing path. 
LRD Trojans add explicit trigger and payload circuitry, but analog sensor signals are continuous, noisy, and highly sensitive to circuit loading. 
Additional circuitry may change path impedance, increase leakage current, or distort normal sensor outputs, while the trigger must still distinguish crafted physical signals from benign environmental variations. \xxx addresses this challenge by separating trigger monitoring from the native sensing path. It uses high-impedance parallel monitoring to observe signal features while minimizing loading effects during normal operation. To improve selectivity, LRD Trojans combine single-condition and multi-condition detection over features such as frequency and amplitude, together with temporal validation. 
Only when the crafted trigger satisfies the predefined rule for a sufficient duration does the Trojan activate the payload.


For PED Trojans, the main challenge is inducing exploitable behavior without explicit trigger logic. Dedicated detection logic and payload modules increase hardware footprint~\cite{bhunia2014hardware, tehranipoor2010survey}, especially in tightly constrained sensor circuits. PED Trojans therefore avoid explicit monitoring circuitry and instead rely on native analog effects in sensor components. However, without a separate trigger decoder, the Trojan must still transform crafted physical inputs into targeted sensor output corruption while preserving normal behavior under benign inputs. 
\xxx addresses this challenge by systematically analyzing sensor modules and circuit characteristics to identify parameters that can be tuned without noticeably degrading normal operation. 
By tuning effects such as nonlinearity, rectification, and frequency-dependent coupling, PED Trojans convert crafted physical inputs into low-frequency output deviations. This allows SenTrojan to induce exploitable behavior in sensors resistant to known signal injection attacks, or amplify weak existing vulnerabilities with minimal component insertion.


Our contributions are summarized as follows:
\begin{itemize}[leftmargin=10pt,nosep]
\item We present the first systematic study of analog sensor Trojans, where crafted physical signals activate malicious behavior without software privileges or digital access.
\item We design \xxx, a two-paradigm framework for analog sensor Trojans that combines broadly applicable logic-driven Trojans with minimally inserted effect-driven Trojans.
\item  We implement and evaluate \xxx across 14 sensors spanning 6 transduction principles, demonstrating attack feasibility, stealthiness, and real-world impact.
\end{itemize}

%% file: sections/background.tex
\section{Background}\label{background}
\subsection{Sensor Structure}
\begin{figure}[t]
    \centering
    \includegraphics[width=\linewidth]{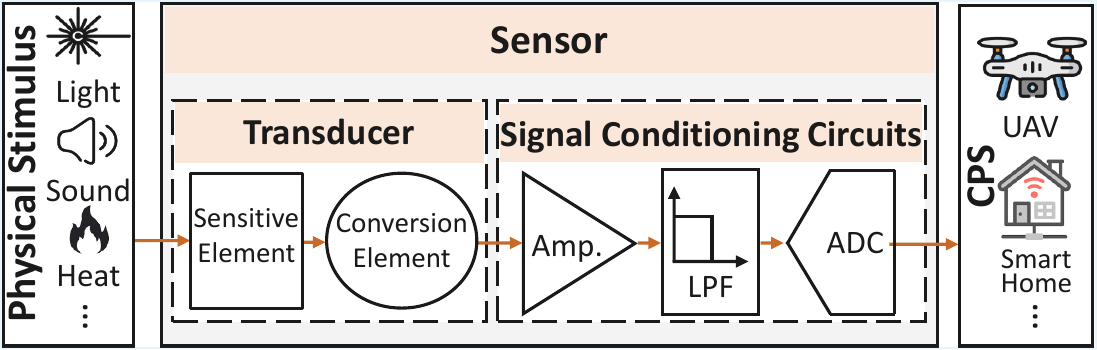}
    \caption{The typical hardware structure of a sensor.}
    \label{fig-sensor_structure}
\end{figure}

A sensor is designed to generate electrical measurements in response to a physical stimulus~\cite{gopel1989sensors}, which is then transmitted to cyber-physical systems (CPS), such as unmanned aerial vehicles and smart homes, as illustrated in Fig.~\ref{fig-sensor_structure}.  A modern sensor comprises four main components: transducer, amplifier, filter, and analog-to-digital converter (ADC).
The transducer is the core component of a sensor, consisting of a sensitive element and a conversion element. The sensitive element detects a specific physical stimulus, while the conversion element translates this stimulus into an analog electrical signal.
Because the transducer’s output is generally weak and vulnerable to noise, the signal conditioning circuit incorporates amplifiers and filters to boost the signal and suppress noise. To make the signal suitable for digital processing, an ADC converts the analog output to digital form.

\subsection{Hardware Trojan}
Hardware Trojans are malicious circuits covertly embedded into hardware systems, such as additional or altered logic, that typically remain inactive until triggered by rare or specific conditions. The common methods to activate Trojans are specific digital inputs or software commands. Once activated, they can compromise security by leaking sensitive data or degrading system reliability. These Trojans pose a significant threat, particularly in complex, multi-party semiconductor supply chains, because they can be inserted at various stages of the hardware lifecycle, including design, fabrication, and assembly ~\cite{puschner2023red, miller2013supply}.

The design of a hardware Trojan can be decoupled into two parts: \textit{trigger} and \textit{payload}.
The trigger determines under which conditions the malicious circuit becomes active, while the payload decides which malicious functionality the Trojan can achieve.
Even though the trigger does not achieve any malicious functionality, it can largely improve the stealthiness of a  hardware Trojan.
An ideal trigger condition is nearly impossible to meet in the normal operation of the hardware, e.g., a random 1024-bit digital input.

%% file: sections/threatmodel.tex
\section{Threat Model}

\subsection{Attacker Model}


We consider an adversary who can compromise sensors before they are deployed or during later maintenance and assembly.
This capability is consistent with modern sensor supply chains as shown in Appendix~\ref{app-lifecycle}, where design, fabrication, packaging, PCB assembly, and maintenance may be performed by different parties across organizational or geographic boundaries.
\rev{Specifically, Trojans may be inserted at the \textit{design stage}, where a malicious insider modifies circuit parameters or component choices~\cite{zhang2014detrust,waksman2011silencing,trippel2021bomberman,imeson2016non}; the \textit{manufacturing stage}, where an untrusted foundry or fabrication process introduces layout or process-level modifications~\cite{yang2016a2,dharsee2023jinn,de2020hartbleed}; or the \textit{post-manufacturing stage}, where a compromised assembler or maintenance adversary replaces components, adds small circuitry, or alters PCB traces~\cite{shwartz2017shattered,mitre_attack,ico_supply_chain_2023}.
The adversary may be motivated by sabotage, espionage, or economic gain. Our design does not depend on the insertion stage. It assumes only that the adversary can embed a Trojan in the sensor’s analog signal path without disrupting normal validation. Appendix~\ref{app_insert} details the capabilities and limitations at each stage.}

\subsection{Attacker's Capabilities} 

To avoid raising suspicion during inspection or functional validation, the adversary is constrained in both the type and number of inserted components. We assume that the adversary does not add digital modules, such as processors or microcontrollers, which are uncommon in sensor front ends and require additional support circuitry. Instead, the adversary may insert or modify only small-scale integrated (SSI) circuits, e.g., logic gates and analog switches, and discrete components, e.g., resistors, capacitors, and diodes. These components are common in sensor circuits and can be introduced with limited impact on the overall architecture.




\subsection{Attack Goal}\label{attack_goal}

The adversary aims to compromise sensor measurements by activating the implanted Trojan with a crafted physical signal. We consider three representative attack goals.

\begin{itemize}[leftmargin=10pt,nosep]
    \item \textbf{Static spoofing.} 
    The Trojan forces the sensor to output a constant, predefined erroneous value. This effect can persist after activation without continuous intervention and may require a system reset or sensor replacement for recovery.

\item \textbf{Dynamic spoofing.} 
The Trojan manipulates sensor readings into attacker-controlled time-varying signals. Unlike static spoofing, dynamic spoofing can produce measurements that may appear plausible to downstream systems.

\item \textbf{Permanent denial-of-service (PDoS).}
The Trojan renders the sensor unable to provide valid measurements. This effect may result from permanent physical damage or from repeatedly forcing the sensor into a protection state.

\end{itemize}




\begin{figure*}[t]
    \centering
    \includegraphics[width=1.0\linewidth]{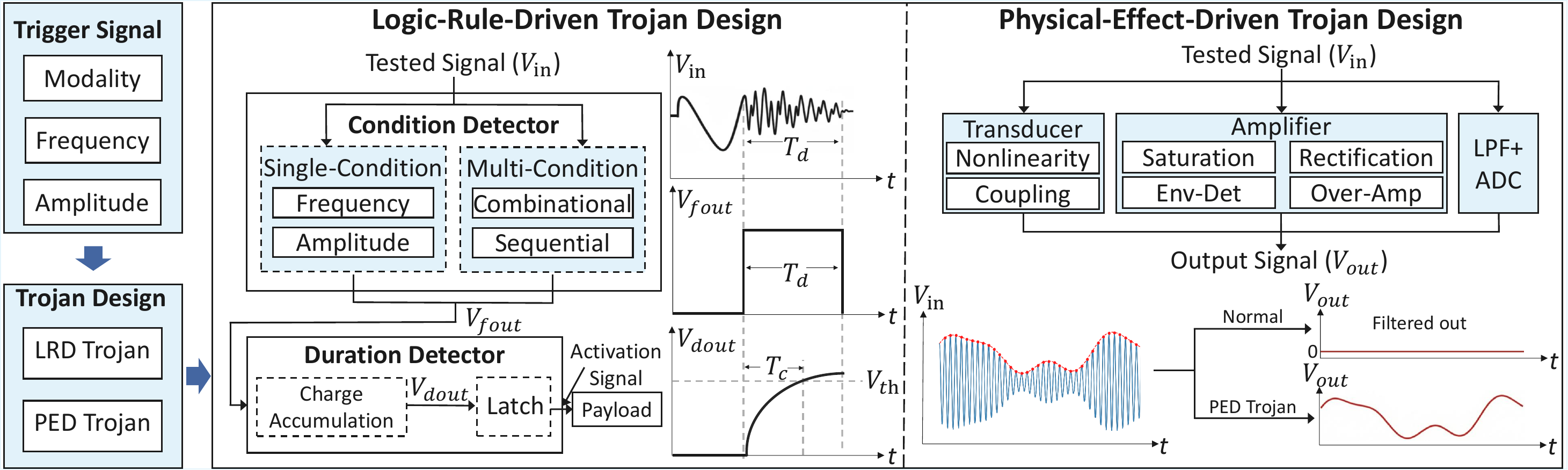}
\caption{An illustration of the \xxx design. Left: Trigger signal specification. Center: Logic-Rule-Driven (LRD) Trojans utilize discrete logic to validate signal features and duration ($T_d$) before latching the activation signal. Right: Physical-Effect-Driven (PED) Trojans leverage component physics like rectification to extract baseband payloads from high-frequency inputs, effectively bypassing protection filters to manipulate sensor measurements.}
    \label{fig_overview}
\end{figure*}

\subsection{Design Requirement}\label{design_requirement}
We consider that a successful \xxx attack should satisfy the following requirements:
\begin{itemize}[leftmargin=10pt,nosep]
  \item \textbf{Trojan functionality.} 
  The Trojan should be reliably activated by a predefined physical signal and induce the intended sensor malfunction after activation.
  \item \textbf{Trojan stealthiness.} 
  The Trojan should preserve normal sensor behavior before activation and avoid false activation under benign environmental stimuli, even under practical functional validation~\cite{yang2016a2}.

  \item \textbf{Minimum components.}
The Trojan should use as few inserted or modified components as possible to reduce hardware overhead and signs of tampering.
\end{itemize}

%% file: sections/design.tex
\section{Design} \label{Design}
\rev{This section details the \xxx design illustrated in~\cref{fig_overview}. Because attack execution requires the signal modality and trigger parameters to be tailored to each target sensor (\cref{design_trigger_signal_design}), we develop a Trojan framework with two analog-hardware activation paths: an external logic-rule-driven (LRD) path and an internal physical-effect-driven (PED) path. LRD uses explicit high-impedance monitoring, feature validation, and a separate payload module, enabling flexible and controllable trigger-payload configurations across sensor types at the cost of additional circuitry. In contrast, PED introduces or tunes analog physical effects, such as envelope detection, to minimize component overhead and enable stealthy dynamic spoofing. We detail the LRD and PED mechanisms and their circuit designs in~\cref{sec:basic,sec:sensor}, with emphasis on attack effectiveness and stealth.}

\subsection{Trigger Signal Design}\label{design_trigger_signal_design}
Trigger signal architecture necessitates precise modality selection and parameter specification.

\textbf{Modality selection.} Prioritizing practical long-range injection over close contact methods suggests four optimal modalities: laser, sound, ultrasound, and EMI. For example, using ultrasound to trigger a Trojan is commonly considered more stealthy and practical than applying pressure to a pressure sensor. We exploit native transducer sensitivities to aforementioned signal modalities, such as accelerometer acoustic susceptibility~\cite{trippel2017walnut} or microphone ultrasonic vulnerability~\cite{zhang2017dolphinattack}. Alternatively, we integrate supplementary transducers, like photoresistors for optical responsiveness. A detailed comparison of these signal modalities is available in~ Appendix~\ref{app_triggersignal}.

\textbf{Parameter design.} Parameter optimization targets frequency coupling efficiency to extend attack range, requiring physical testing like sweeping. In terms of amplitude, higher signal levels typically induce stronger interference. The design of the Trojan activation mechanism necessitates a comparative analysis of injected versus legitimate signal characteristics. Specifically, we characterize legitimate signals across their datasheet-specified operational ranges to identify and exploit characteristic disparities as trigger thresholds.

\subsection{Logic-Rule-Driven Trojan Design} \label{sec:basic}
To prevent accuracy degradation, we employ a parallel monitoring circuit that operates in a high-impedance state during quiescence, thereby preserving the native sensor precision with minimal signal perturbation. Furthermore, to evade detection during brief factory functional tests~\cite{mcfadden2010supply}, the Trojan incorporates a temporal activation threshold that exceeds standard durations. LRD Trojan architecture integrates three core modules: the trigger condition detector, trigger duration detector, and payload.

\begin{figure}[t]
    \centering
        \subfigure[Frequency detector. ]{
        \begin{minipage}[t]{0.59\linewidth}
            \centering
            \includegraphics[width=1\textwidth]{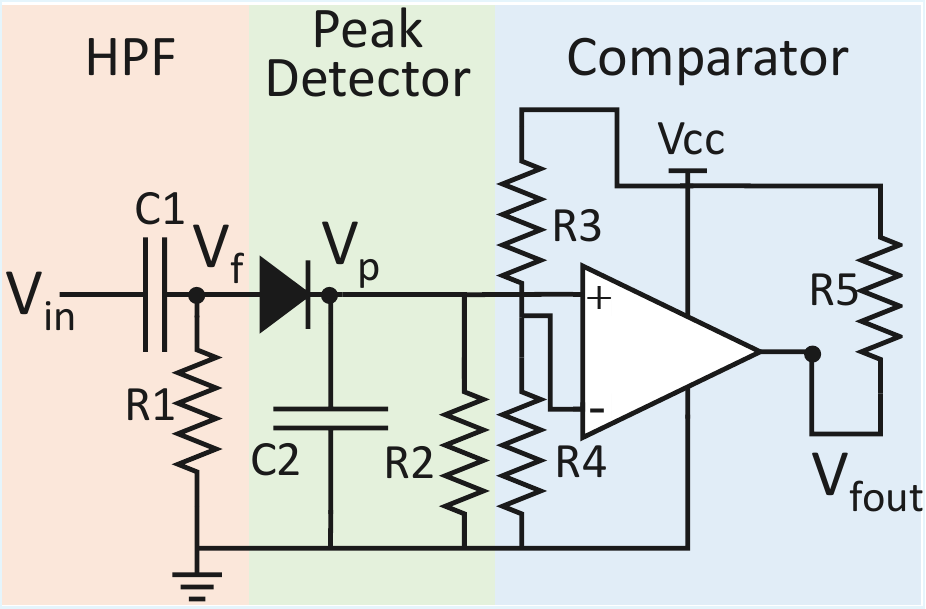}
        \end{minipage}
        \label{fig_design_freq}
    }
    \subfigure[Amplitude detector.]{
        \begin{minipage}[t]{0.3\linewidth}
            \centering
            \includegraphics[width=1\textwidth]{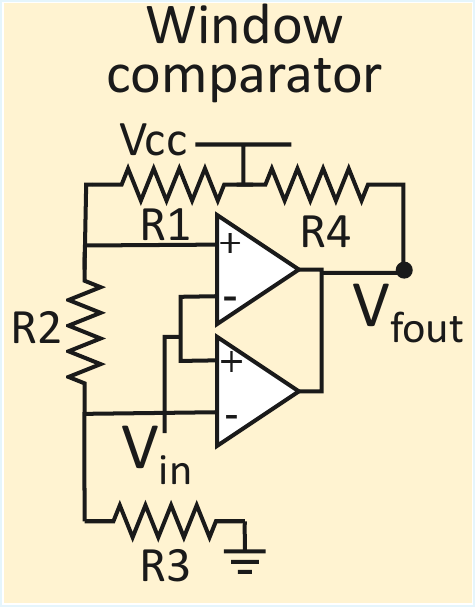}
        \end{minipage}
        \label{fig_design_volt}
    }

    \caption{Schematics of a single condition detector. }
    \label{fig_design_single}
\end{figure}

\begin{figure}[t] 
    \centering
    \subfigure[Waveforms under 2 kHz input.]{
		\begin{minipage}[t]{0.45\linewidth}
			\centering
			\includegraphics[width=1\textwidth]{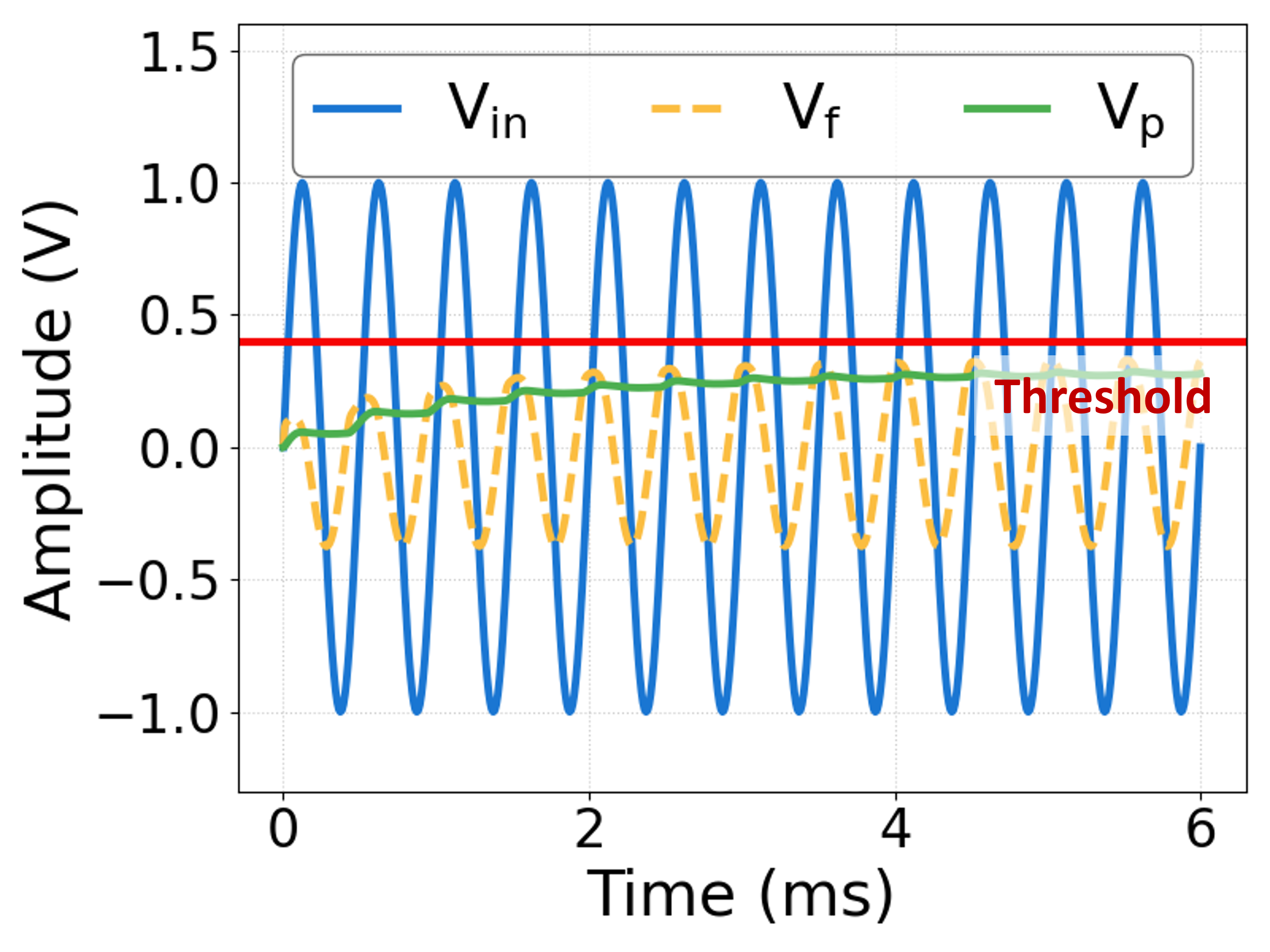}
		\end{minipage}
		\label{fig_frequency_wave2k}
        }
    \subfigure[Waveforms under 20 kHz input.]{
		\begin{minipage}[t]{0.45\linewidth}
			\centering
			\includegraphics[width=1\textwidth]{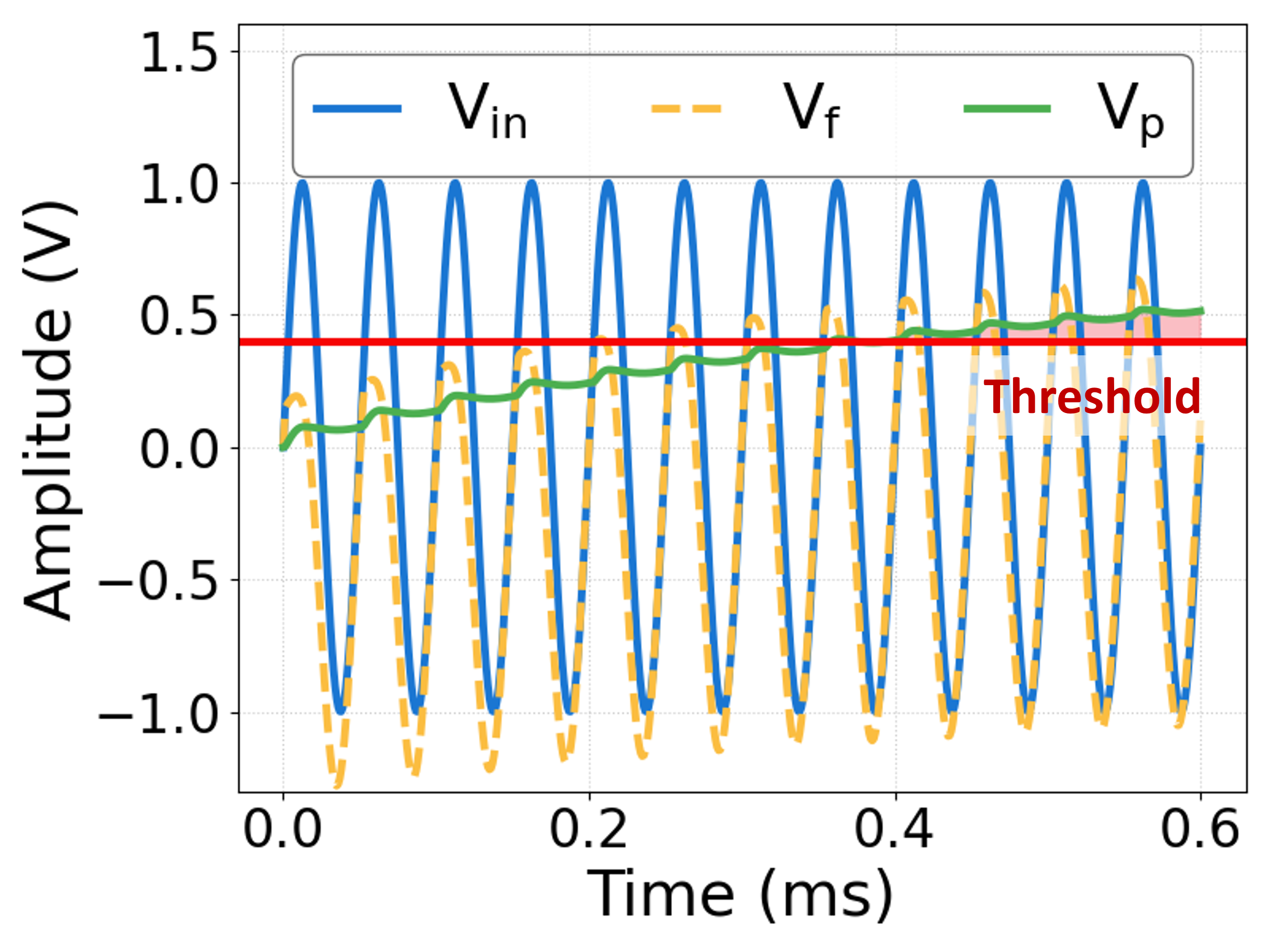}
		\end{minipage}
		\label{fig_frequency_wave20k}
	}
    \caption{Simulation waveforms of the frequency detector. $\mathrm{V_{in}}$ is the input, $\mathrm{V_f}$ is the filtered signal, and $\mathrm{V_p}$ is the peak-detector output. When $\mathrm{V_p}$ exceeds the threshold, the subsequent comparator is activated, triggering the payload.}
    \label{fig_sim_fre}
\end{figure}

\subsubsection{Single-Condition Detector} 
Common characteristics of an analog signal include frequency, amplitude, and waveform~\cite{xiao2025sok}. However, the waveform cannot be efficiently exploited for Trojan triggering, as complex shape discrimination necessitates prohibitive area and power overheads.
As illustrated in~\cref{fig_design_single}, the single-condition circuit translates true detection logic into a high-level signal, denoted as $V_{fout}$, which activates the trigger duration detector.

\textbf{Frequency.}
To maintain high input resistance and preserve monitoring accuracy, our frequency triggering mechanism detects the preset frequency using a cascade of a frequency detector, a peak detector, and a comparator, as depicted in~\cref{fig_design_freq}.
\rev{Frequency detection can use LC resonance (frequency-selective behavior in an inductor-capacitor circuit) or high-pass filters (filters that pass frequencies above a cutoff frequency).}
Crucially, even if environmental noise coincides with the trigger frequency, the subsequent comparator prevents false activation by enforcing a strict amplitude threshold.
Initially, the filter isolates the frequency range, setting cut-off frequencies, $f_p=1/(2\pi R_1C_1)$, above normal sensor operation to mitigate non-ideal characteristics. An enhanced peak detector subsequently minimizes adjacent frequency interference and improves reliability, utilizing diodes for unidirectional current and capacitors as analog memory to retain the peak value. To ensure subsequent circuit stages remain unaffected, resistor $R_2$ facilitates efficient charge discharge, isolating downstream stages, necessitating component values where $1/(2\pi R_2C_2) \approx f_{in}^{'}$ matches the attacker-injected frequency $f_{in}^{'}$.

Reference voltages determined by $R_3$ and $R_4$ necessitate sufficiently large $R_4/(R_3 + R_4)$ ratios to reject noise, exemplified by our $R_4 \cdot V_{cc}/(R_3 + R_4) > 0.2~V$ requirement countering sub-$0.2 V$ noise. Simulation results in~\cref{fig_frequency_wave2k} and~\cref{fig_frequency_wave20k} alongside parameter settings in~\cref{fig_frequency_circuit} demonstrate peak detector output amplitude scaling with input frequency.

Finally, a comparator isolates signals exceeding threshold voltages, mitigating filter imperfections to prevent noise-induced false activation. 
We utilize a $1~k\Omega$ pull-up resistor $R_5$ to enhance load handling capability. The parameter requirement for triggering is: 
\begin{equation}\label{frequency}
V_{in}\sqrt{\frac{R^2_1}{R^2_1+\frac{1}{\omega^2C^2_1} } } -V_{d} > V_{cc}\frac{R_4}{R_3+R_4}
\end{equation}
where $V_{in}$ is the input signal and $V_{d}$ is the diode's forward voltage.

\textbf{Amplitude.}
To ascertain whether a measured voltage falls within a specific threshold range, attackers can employ various circuits such as average detectors, limiters, and comparators.
We utilize a window comparator circuit to exemplify the amplitude detection mechanism, which consists of two comparators as depicted in \cref{fig_design_volt}. This topology guarantees high input resistance and produces a high level activation signal exclusively when the voltage falls within the designated bounds. The comparator configuration relies on the amplitude relationship between the injected and legitimate signals.  For instance, an amplitude threshold configuration is required when the injected signal exceeds the legitimate signal's normal operational range.
The window comparator output is defined as:
    \begin{equation}\label{Comparator}
        V_{fout} =
            \begin{cases}
                Vcc & \text{if } V_{{in}} \in [V_- , V_+] \\
                0 & \text{if } V_{{in}} \notin [V_- , V_+]
            \end{cases}
    \end{equation}
where $ V_{fout}$ is the comparator output, $ V_{in}$ is the sensor voltage, and $ V_-$, $ V_+$ represent the lower and upper reference voltages, respectively.
Balancing power consumption and interference immunity mandates mega-ohm range level selection for $R_1$, $R_2$, and $R_3$, while pull-up resistor $R_4$ stabilizes the operational amplifier output by enhancing load handling and preventing undefined floating states.

AC signal amplitude detection requires diode rectification and filtering to yield smooth DC signals, then processed via DC trigger logic to ascertain input amplitudes.
\subsubsection{Multi-Condition Detector}
To improve robustness and suppress false triggers, the detector requires multiple features of the same input signal, such as amplitude and frequency, to satisfy predefined conditions. In the combinational mode, feature detectors operate in parallel, and the trigger is asserted only when all detector outputs are valid within a specified time window. In the sequential mode, analog switches control the activation of successive detectors: each detector is enabled only after the preceding feature has been detected, thereby enforcing a prescribed temporal order. In both modes, circuits consisting of diodes, capacitors, and resistors retain each detection result for a configurable duration. If a required feature is missing or arrives outside its allowed window, activation is blocked; during sequential detection, the process returns to its initial stage. The payload is activated only after the complete feature set has been verified with the required timing.


\subsubsection{Duration Detector}

\begin{figure}[t]
    \centering
    \includegraphics[width=0.98\linewidth]{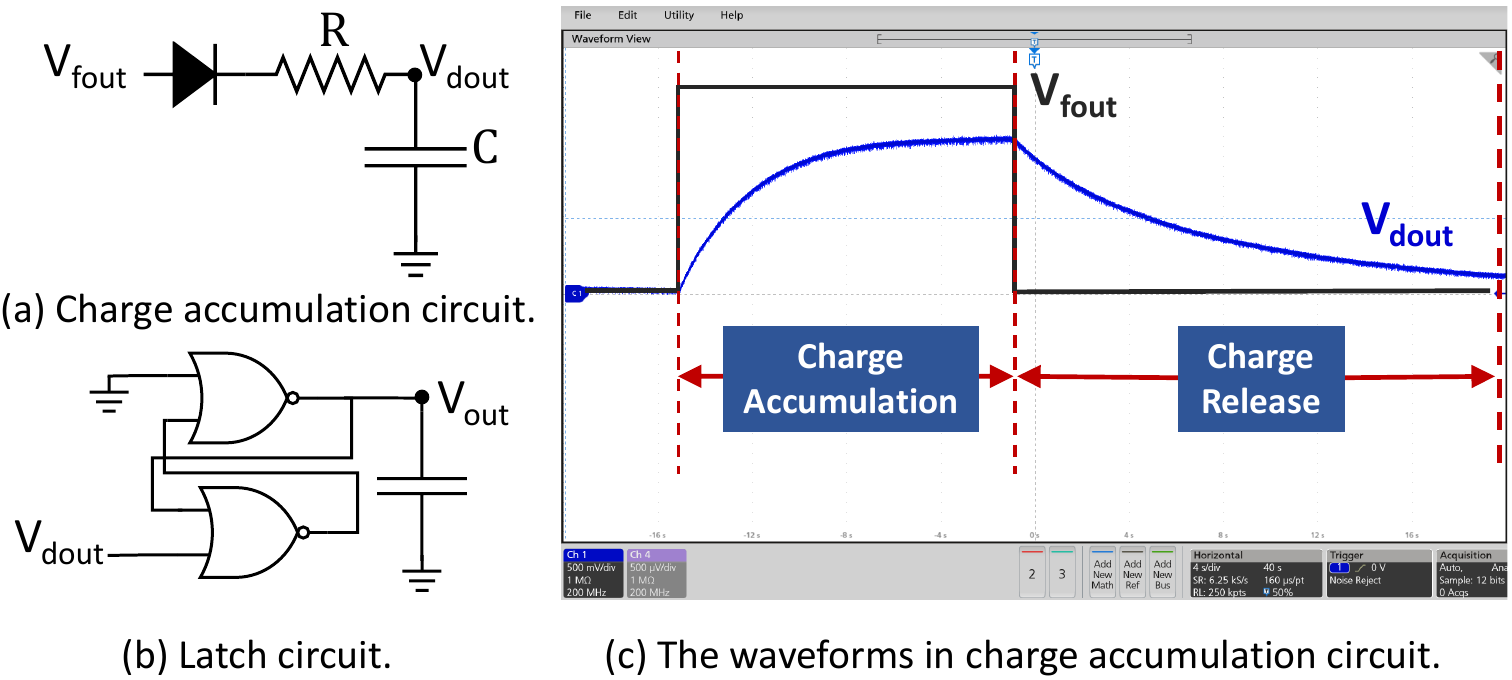}
    \caption{ The duration detector schematics and the simulation waveforms.}
    \label{fig_design_duration}
\end{figure}    

The duration detector can be realized using a linear integrator circuit, a ramp generator, or similar topologies.
We adapt the concept of counter-based triggers prevalent in digital circuits~\cite{waksman2011silencing}, by leveraging charge accumulation in capacitors coupled with a latch to achieve similar effects, as depicted in~\cref{fig_design_duration}.
We omit duration detectors for extra transducer Trojans to minimize component counts, since manufacturers and users commonly restrict testing for their intended sensing modalities in common practice.

\textbf{Charge accumulation circuit.} 
True trigger condition logic initiates capacitor charging, defining trigger duration as the voltage rise interval to threshold $V_{th}$. The minimum voltage of logic high serves as the threshold voltage, eliminating the need for extra comparators. Consequently, injection duration $T_c$ can be calculated as:

\begin{equation}
    T_c = RC({\ln{V_{fout}}-\ln({V_{fout}-V_{th}}}))
    \label{eq_charge_time}
\end{equation}
where $V_{fout}$ is the desired output of the trigger condition detector and $R$ is the series resistance. 

In practical implementations, capacitors exhibit natural leakage due to material imperfections and ambient conditions. We leverage this inherent discharge mechanism to reset the state without additional components, preventing charge accumulation from compromising the trigger duration detector's accuracy.
Physical setup and validation results are presented in~\cref{fig_design_duration}. Unless otherwise specified, we utilize $R = 470~k\Omega$ and $C = 10~\mu F$, resulting in a 4-second rise time, which is sufficient to evade factory testing yet enable efficient attacks.


\textbf{Latch.}
When exploitable signal nodes are limited to sensor outputs, activation can disrupt the triggering condition. We therefore employ latch mechanisms to preserve the activated state independently of subsequent signal variations. Persistent effects demand strict logic low initialization achieved via a capacitor-based power-on-reset design, where capacitors ensure reliable initialization with negligible power overhead.


\subsubsection{Payload}\label{sec-parallel}

Payload implementation can utilize CMOS transmission gates, BJT switches, or MEMS relays to modify sensor outputs. We employ analog switches to toggle transmission paths and mimic broken or shorted states as required to force the sensor's output to switch to a preset voltage signal for static spoofing. 
Alternatively, the latch-up mechanism induces direct supply-to-ground shorts, leading to PDoS. In devices lacking protection mechanisms, the resulting overcurrent causes permanent physical damage, such as trace burnout or component degradation, rendering the sensor permanently inoperable. In contrast, although modern devices equipped with voltage regulators often incorporate thermal shutdown or current limiting, the Trojan can exploit these safeguards by persistently re-triggering the fault. This forces the sensor into a continuous shutdown loop, indirectly but effectively achieving a PDoS condition. Finally, dynamic spoofing mandates extra transducers to modulate sensor output, enabling real-time control of injected values to realize complex time-varying waveforms.

\begin{tcolorbox}[
  colback=gray!10,
  colframe=blue!70!black,
  boxrule=0pt,
  leftrule=1.5mm,
  arc=0pt,
  outer arc=0pt,
  left=6pt,
  right=6pt,
  top=4pt,
  bottom=4pt
]
\textcolor{green!60!black}{\textbf{Insight 1}}: LRD Trojans trade structural stealth for controllability, using explicit trigger and payload circuits to enable reliable triggering and diverse attack effects across sensor types.
\end{tcolorbox}


\subsection{Physical-Effect-Driven Trojan Design} \label{sec:sensor}
While LRD Trojans enable diverse attacks, they prove complex and resource-intensive. Leveraging internal physical effects, we design stealth-oriented mechanisms with lower hardware costs. 
Avoiding explicit trigger logic, our approach emulates analog circuit characteristics via subtle parameter modifications or minimal additions, eliminating explicit activation circuits to enhance Trojan stealthiness. Utilizing components within standard manufacturing tolerances causes forensic analysts to classify Trojans as manufacturing defects or cost-cutting measures rather than malicious components, making them harder to detect during standard inspection.

Injected signals target frequency bands beyond LPF cutoff frequencies, such as resonant frequencies, yet face attenuation from sensor filters and protective circuitry. To bypass defenses, our Trojan translates high-frequency signals into low-frequency ones while preserving payload information, detailed in~\cref{fig_overview}.  Unlike LRD Trojans, this conversion typically relies on the sensor's existing coupling path rather than an additional trigger circuit.  We therefore systematically analyze each sensor component and identify exploitable analog effects that can be induced or amplified with minimal hardware changes.



\subsubsection{Amplifier}
 We propose four methods for inserting Trojans into amplifiers: saturation, envelope detection, rectification, and over-amplification, corresponding to enhancing nonlinearity, inducing nonlinearity, and using a nonlinearity-free approach. 


\begin{figure*}[t]
    \centering
    \subfigure[The saturation circuit and the simulation waveforms.]{
        \begin{minipage}[t]{0.45\linewidth}
            \centering
            \includegraphics[width=1\textwidth]{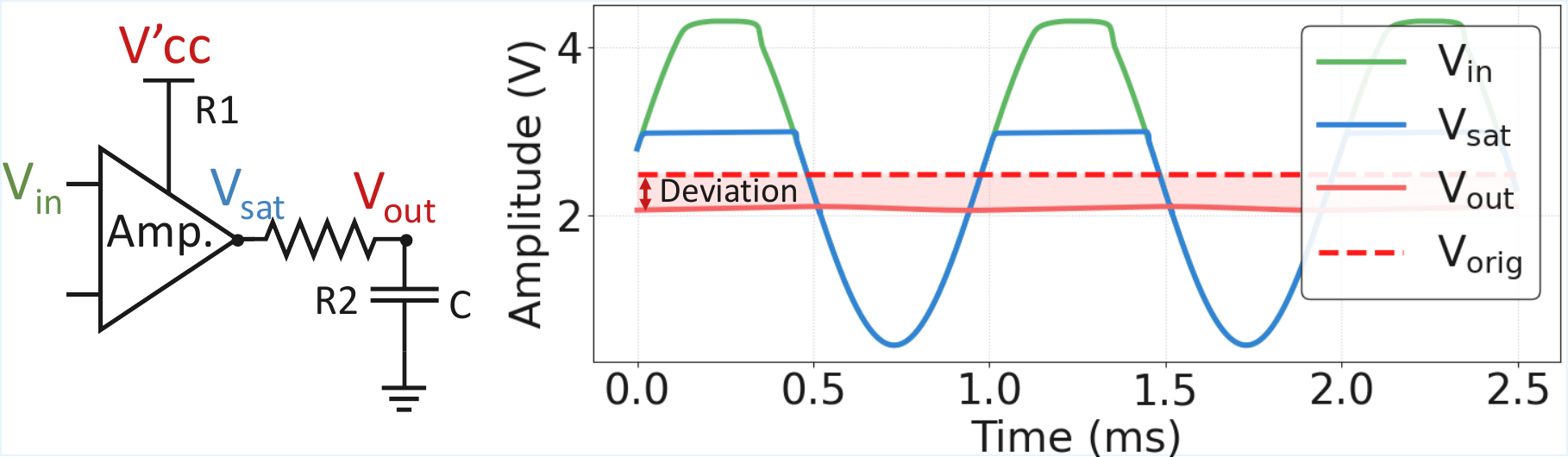}
        \end{minipage}
        \label{fig_amp1}
     }
    \subfigure[The rectification circuit and the tested waveforms.]{
        \begin{minipage}[t]{0.45\linewidth}
            \centering
            \includegraphics[width=1\textwidth]{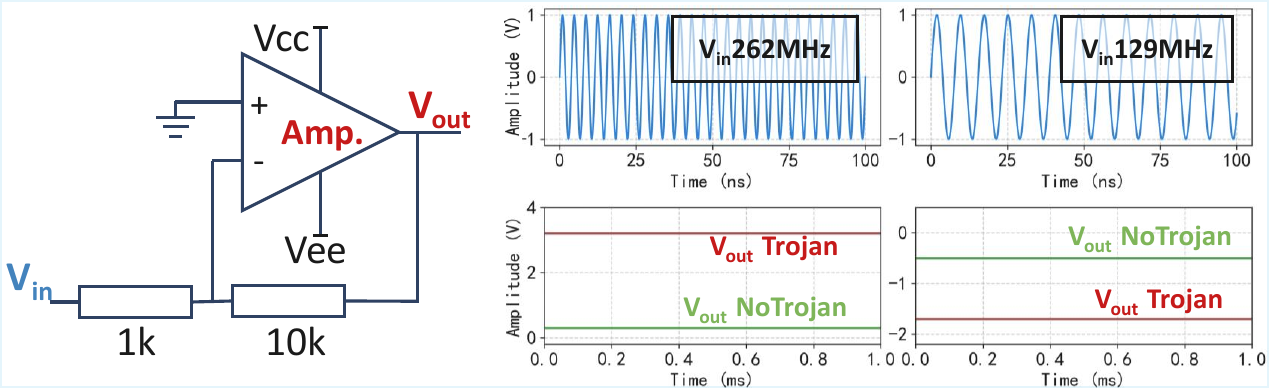}
        \end{minipage}
        \label{fig_amp0}
     }
    \subfigure[The envelope detection circuit and the simulation waveforms.]{
        \begin{minipage}[t]{0.45\linewidth}
            \centering
            \includegraphics[width=1\textwidth]{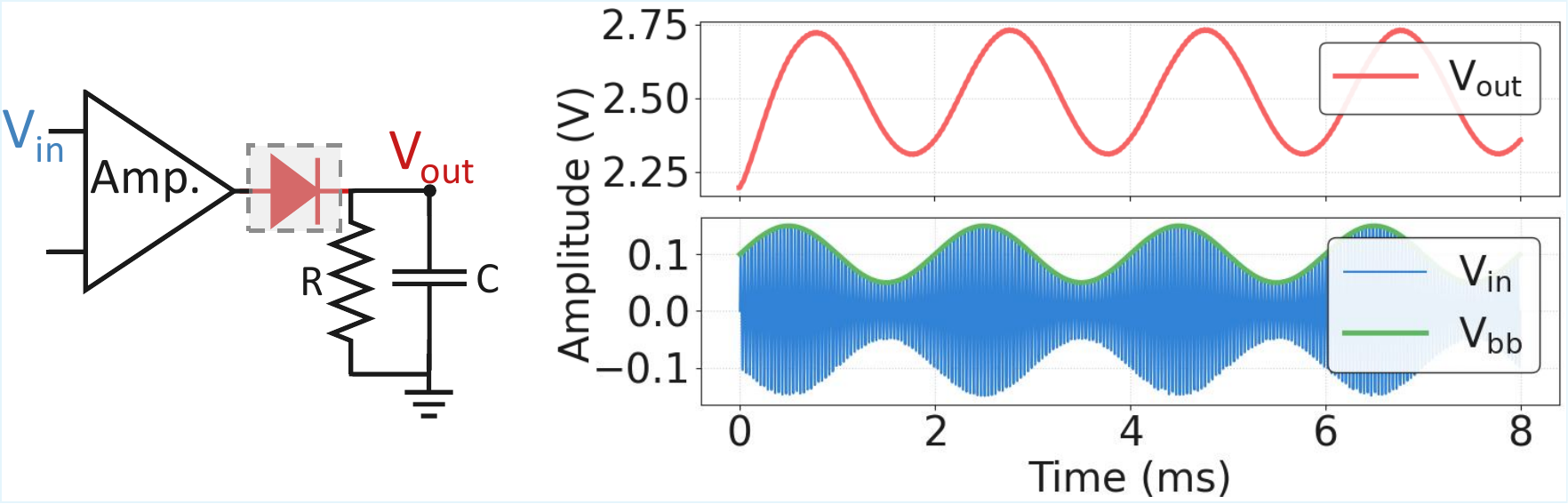}
        \end{minipage}
        \label{fig_amp2}
    }
    \subfigure[The over-amplification circuit and the simulation waveforms.]{
        \begin{minipage}[t]{0.45\linewidth}
            \centering
                \includegraphics[width=1\textwidth]{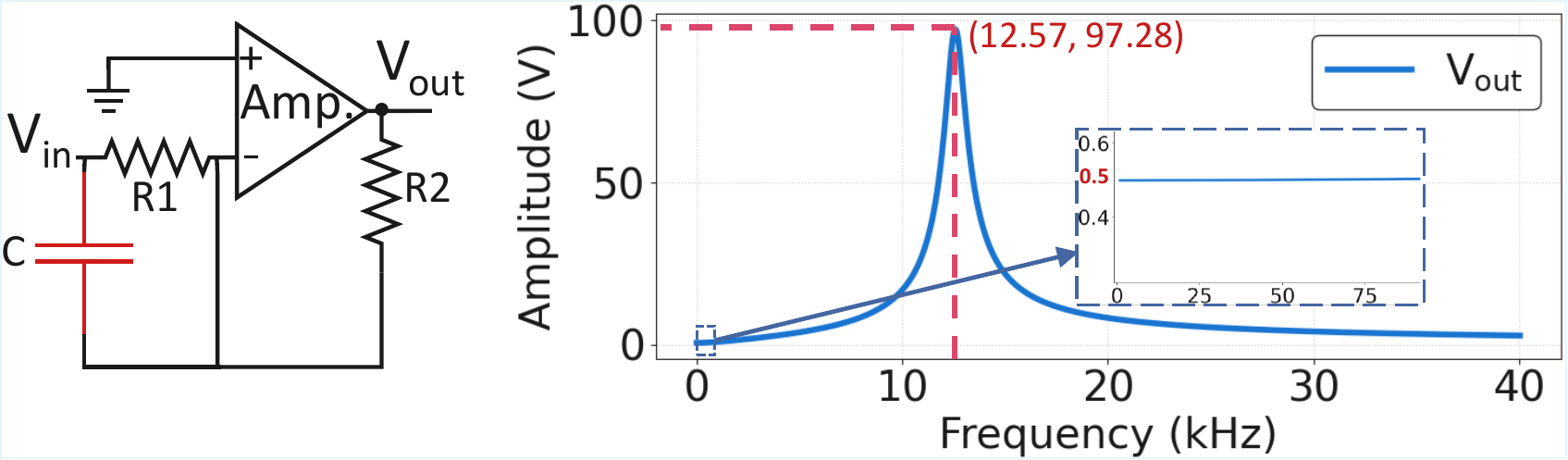}
        \end{minipage}
        \label{fig_amp3}
    }
    \caption{The circuit schematics and experimental results of the PED Trojans in the amplifier. }
\end{figure*}

\textbf{Saturation.}
Saturation-based Trojans exploit amplifier amplitude limits for baseband signal recovery. We analyze saturation phenomena in Appendix~\ref{appen-satu} to identify exploitable methods. Since symmetrical saturation clipping preserves the DC baseline, we propose asymmetric cutoff techniques introducing DC offsets when injected AC signals exceed dynamic range.

Three methods induce asymmetric clipping. First, modifying DC bias resistors shifts the static operating point from $V_{CC}/2$, altering the positive and negative range to cause asymmetric saturation. Second, selecting amplifiers with lower $V_{max}$ restricts output limits despite identical power supplies. Third, inserting resistors between the main $V_{CC}$ and the amplifier power port reduces the supply voltage to $V_{cc}^{\prime}$ above normal sensor output voltages to maintain operation. This induces voltage drops, lowering saturation thresholds without affecting normal operation.

With the injected AC signal included in \(V_{\mathrm{in}}\), the amplifier output is clipped and exhibits a DC offset. Filtered amplitude $V_{out}$ follows:
\begin{equation}
    \rev{V_{out}(t) = \frac{1}{T_d}\int_{t}^{t+T_d} V_{aout}(\tau)\,\mathrm{d}\tau.}
\end{equation}
\rev{where $T_d$ denotes the injection signal duration and $V_{aout}(\tau)$ is the amplifier output. Theoretical maximum offsets $(V_{max}-V_{max}^{\prime})/2$ enable AM signals to spoof any waveform within this range when the resonant period $T$ satisfies $T \ll T_d$}. Saturation Trojans require legitimate signal amplitudes below injection levels to ensure proper sensor operation. Simulations with the parameters in Appendix~\ref{sec_sim_amp1} yield results in~\cref{fig_amp1}, confirming that input amplitudes exceeding $V_{max}^{\prime}$ undergo clipping, inducing output deviations.



\textbf{Rectification.}
The nonlinear rectification effect in amplifiers can convert high-frequency AC signals into DC offsets. This phenomenon stems from the inherent nonlinearity of the input differential pair's current-voltage characteristics. However, the susceptibility to this rectification varies markedly across different amplifier topologies. Consequently, adversaries can strategically exploit this variability by targeting amplifiers with pronounced nonlinear sensitivity to maximize the efficacy of physical signal injection attacks. The resulting attack impact is illustrated in~\cref{fig_amp0}.

\textbf{Envelope detection.}
Beyond enhancing existing nonlinearity, inducing new nonlinearity enables Trojan insertion.
For nonlinear components, a diode is preferred first as it is commonly used in sensors to block reverse current without disrupting normal circuit behavior.  
The diode's forward conduction does not impact the circuit's operation, as it works with the DC bias configuration of operational amplifiers.
  
To enable targeted frequency signal demodulation, we design a circuit comprising a diode, capacitors, and resistors, allowing Trojan insertion via RC parameter tuning for envelope detection. Furthermore, absent physical diodes, amplifiers enforce unidirectional flow mimicking diode behavior to extract baseband signals, albeit within restricted amplitude range.
Carrier frequency dictates injection efficiency, requiring a balance among sensor frequency response and baseband signal bandwidth, while determining demodulation parameters and attack effectiveness.
Diode conduction during positive high-frequency cycles rapidly charges capacitors via low forward resistance. Signals dropping below the capacitor voltage trigger reverse bias, causing cutoff and slow discharge through load resistors. Brief conduction combined with discharge durations exceeding signal periods enables AM waveform envelope approximation. Diodes mimic open circuits for low-frequency legitimate signals, preserving quality.

Simulations involving circuits with and without diodes confirm successful baseband signal restoration for both Trojan methods, utilizing schematics in~\cref{fig_amp2} and parameters in Appendix~\ref{sec_sim_amp2}. Modifying RC-tuned envelope detectors or adding diodes enables hidden signal demodulation from high-frequency carriers without disrupting normal operation. Envelope detection implementation necessitates sensor tolerance for small voltage drops, as added diodes introduce slight voltage reductions.

\textbf{Over-amplification.}
Beyond nonlinearity, attackers induce overamplification via frequency response modification, using high-frequency triggers to drive sensors into saturation and cause DoS. 
As temperature and pressure sensors operate in low-frequency bands, we ensure stealth by selecting triggers outside this legitimate bandwidth. Inserting a capacitor $C$ parallel to the input resistor $R_1$ leverages capacitive bypassing at high frequencies, yielding gain as: 
\begin{equation}
    \frac{V_{out}}{V_{in}} = \frac{R_2}{R_1 \parallel \frac{1}{j\omega C}}
\end{equation}
where $V_{in}$ and $V_{out}$ denote signals and $R_1$, $R_2$, and $C$ values corresponding to the circuit components. 

Signal passage at the preset frequency sharply reduces the combined $C$ and $R_1$ impedance, increasing gain. We determine $C$ via the desired amplification factor, simulating the circuit in \cref{fig_amp3} with parameters in Appendix~\ref{sec_sim_amp3}. Results show 12.57 kHz inputs cause operational amplifier amplification nearly 200 times that of legitimate signals, creating interference, while low-frequency inputs remain accurate. Adding a small resistor mitigates high-frequency environmental noise triggering to further enhance stealth.

\subsubsection{Transducer}
The transduction process is inherently nonlinear and can be formulated as:
\begin{equation}
V_{out}(t)= A_1 V_{in}(t) + A_2 V^{2}_{in}(t) + ... +  A_k V^{k}_{in}(t)
	\label{eq_nonlinarity}
\end{equation}
where  $A_1$ is the gain for the input signal and $A_k$ is the gain of the kth term $V^k_{in}$. 

Nonlinearity commonly arises from manufacturing imperfections and intrinsic material properties. Components designed for the same function can vary significantly in their degree of nonlinearity, especially in the $A_2$ coefficient. Attackers can exploit this variation by selecting components with a higher $A_2$  to insert a Trojan. To activate the Trojan, amplitude modulation is used to restore the baseband signal via nonlinearity. More sensitive transducers can further enhance the attack efficiency.

\subsubsection{LPF and ADC}
Ideal LPFs block all signals above their cutoff frequency.  However, real filters only attenuate these signals.
Attackers can increase the cutoff frequency by modifying the resistance or capacitance. This change reduces the attenuation of a high-frequency attack signal, allowing it to pass through the filter.
To prevent aliasing, the ADC's sampling rate must be at least twice the LPF's cutoff frequency according to the Nyquist criterion. The attack works by injecting a signal whose frequency is higher than half the sampling rate. 
This aliasing causes the ADC to misinterpret the high-frequency attack signal as legitimate, low-frequency sensor data. 

\subsubsection{Payload Design} 
PED Trojans can be designed without a distinct payload module. The trigger signal is crafted to be demodulated or amplified by the circuit itself, which in turn executes a DoS or spoofing attack. This approach presents a trade-off: gaining stealth by eliminating a payload component at the cost of reduced attack generality.


\begin{tcolorbox}[
  colback=gray!10,
  colframe=blue!70!black,
  boxrule=0pt,
  leftrule=1.5mm,
  arc=0pt,
  outer arc=0pt,
  left=6pt,
  right=6pt,
  top=4pt,
  bottom=4pt
]
\textcolor{green!60!black}{\textbf{Insight 2}}: PED Trojans tune native analog effects to induce or amplify exploitable vulnerabilities, improving structural stealth without explicit trigger logic.
\end{tcolorbox}


%% file: sections/Evaluation.tex
\section{Evaluation} \label{sec:case}
\begin{table}[]
\setlength{\tabcolsep}{5pt}
\setlength{\dashlinedash}{1pt}  
\setlength{\dashlinegap}{1.5pt}
\caption{Comparison of direct sensor attacks and Trojans.}
\resizebox{\linewidth}{!}{%
\begin{tabular}{ccccc}
\toprule
\textbf{Sensor} & \textbf{Obj.} & \textbf{Direct}                                           & \textbf{Trojan}                                             & \textbf{Impact}                                             \\ \midrule
Microphone            & \textbf{1}  & \begin{tabular}[c]{@{}c@{}}Range \\ = 0.5~m\end{tabular}     & \begin{tabular}[c]{@{}c@{}}Range \\ = 5.1~m\end{tabular}       & \begin{tabular}[c]{@{}c@{}}Range\\ +920\%\end{tabular}      \\ \hdashline
Temperature           & \textbf{1}    & \begin{tabular}[c]{@{}c@{}}-2.2$\sim$\\ +5.6~°C\end{tabular} & \begin{tabular}[c]{@{}c@{}}-11.6$\sim$\\ +27.1~°C\end{tabular} & 4.9$\times$ Gain                                            \\ \hdashline
Microphone            & \textbf{2}    & Noise                                                       & \begin{tabular}[c]{@{}c@{}}Valid\\ Command\end{tabular}       & \begin{tabular}[c]{@{}c@{}}Command\\ Injection\end{tabular} \\ \hdashline
Inclination         & \textbf{2}    & \begin{tabular}[c]{@{}c@{}}Correct\\ Reading\end{tabular}   & \begin{tabular}[c]{@{}c@{}}159$^\circ$\\ Offset\end{tabular}  & $\Delta$ = 159$^\circ$                                      \\ \hdashline
Light           & \textbf{3}  & \begin{tabular}[c]{@{}c@{}}Linear\\ Output\end{tabular}     & \begin{tabular}[c]{@{}c@{}}Preset\\ Constant\end{tabular}     & \begin{tabular}[c]{@{}c@{}}Logic\\ Inversion\end{tabular}   \\ \hdashline
Infrared        & \textbf{3}  & \begin{tabular}[c]{@{}c@{}}Correct\\ Reading\end{tabular}   & \begin{tabular}[c]{@{}c@{}}Preset\\ Constant\end{tabular}     & \begin{tabular}[c]{@{}c@{}}False\\ Alarm\end{tabular}       \\ \hdashline
Pressure        & \textbf{3}  & \begin{tabular}[c]{@{}c@{}}Correct\\ Reading\end{tabular}   & \begin{tabular}[c]{@{}c@{}}Preset\\ Constant\end{tabular}     & \begin{tabular}[c]{@{}c@{}}Control\\ Loss\end{tabular}      \\ \hdashline
Temperature           & \textbf{3}  & \begin{tabular}[c]{@{}c@{}}Correct\\ Reading\end{tabular}   & \begin{tabular}[c]{@{}c@{}}Preset\\ Constant\end{tabular}     & \begin{tabular}[c]{@{}c@{}}Loop\\ Hijack\end{tabular}       \\ \bottomrule
\end{tabular}
}
\begin{tablenotes}
    \footnotesize
    \item Obj. : Objective; 1: Efficacy Enhancement; 2: Feasibility Expansion; 3: Functional Extension.
\end{tablenotes}
\label{tab-comparion}
\end{table}
To comprehensively evaluate \xxx, we conduct extensive real-world experiments across multiple sensor types and attack scenarios. We begin with an overview of our experimental methodology and key findings in~\cref{sec:eval-overview}, followed by detailed experimental setups and evaluation metrics in~\cref{sec:eval-setup}. 
We then analyze the Trojan efficacy and hardware overhead of \xxx.

\subsection{Overview} \label{sec:eval-overview}
We evaluate a total of 14 sensors across 6 categories to demonstrate the broad applicability of \xxx, as detailed in~\cref{tab:my-table}. Our evaluation reveals consistently high successful trigger rates (82\%-100\%), excellent stealthiness (similarity \textgreater 97\%), and zero false trigger rates for LRD Trojans across all tested sensor types. 
\rev{All similarity errors are below 5\%, which is comparable to the maximum specified measurement error of representative commercial sensors~\cite{NXP_MPX5010,Amphenol_NPC300,Honeywell_MidasS2_H2,Amphenol_A1737,ADI_ADXL321,Infineon_IM72D128V01}, which indicates that the Trojan-induced deviations are unlikely to be distinguishable through routine functional testing.}
Furthermore, we evaluate common hardware-overhead metrics to quantify the performance impact of \xxx. Results show that Trojan-inserted sensors exhibit less than 5\% deviation from benign sensors in time drift, noise floor, and response time, indicating minimal impact on normal sensor operation.

\rev{Our Trojan design process begins by examining the transducer type and determining whether the target sensor can receive remotely injected out-of-band signals. If not, an additional transducer can be introduced. We then characterize the injected signal feature, i.e., amplitude, frequency, or their combination, and implement a feature extraction circuit for reliable triggering. We modified one representative sensor for each case, and all implemented designs are included in the reported results.}


\textbf{Key results summary.}
Our evaluation demonstrates significant improvements over prior sensor attacks. A comprehensive comparison is summarized in \cref{tab-comparion} and detailed below:

\begin{itemize}[leftmargin=10pt, nosep]
\item \textbf{Light sensors:} Unlike traditional laser attacks that mainly amplify sensor readings, our Trojans can force persistently low outputs, potentially disrupting greenhouse lighting control and reducing crop quality~\cite{ MSUPhotoperiod2002,UMassPhotoperiod2017,WiscSupplementalLighting2026}. 


\item \textbf{Microphones:} We extended the DoS range from 0.5 meters to 5.1 meters and achieved a 100\% spoofing success rate, unlike prior work yielding only noise under identical methods, enabling the manipulation of vehicular voice assistants to unlock doors or control smart home systems.

\item \textbf{Infrared sensors:} We spoof infrared sensors previously unexplored to enable suppression and false alarm attacks against intrusion detection systems, achieving 100\% similarity without unintended triggers.

\item \textbf{Pressure sensors:} We spoof brake pressure sensors previously unexplored to induce critical braking failure, causing the simulated vehicle to collide with obstacles.

\item \textbf{Inclination sensors:} We induced sensor-level angle deviations of up to 159°, while the benign sensor remained stable under the same attack setup. In STK simulations, a 30° Trojan-induced attitude deviation caused a right roll and subsequent altitude loss.
\item \textbf{Temperature sensors:} We amplified temperature deviations by 4.9 times compared to physical signal injection alone, extending the range from negative 11.6~°C to positive 27.1~°C, potentially causing chemical reactor runaway and industrial boiler explosions.
\end{itemize}

\subsection{Evaluation Setup} \label{sec:eval-setup}

\subsubsection{Hardware Configuration}

The experimental setup comprises four components: trigger signal generation, attack devices, Trojan implementation, and sensor measurement.

\begin{figure}[t]
    \centering
    \subfigure[Low-frequency signal emission.]{
        \begin{minipage}[t]{0.45\linewidth}
            \centering
            \includegraphics[width=1.0\linewidth]{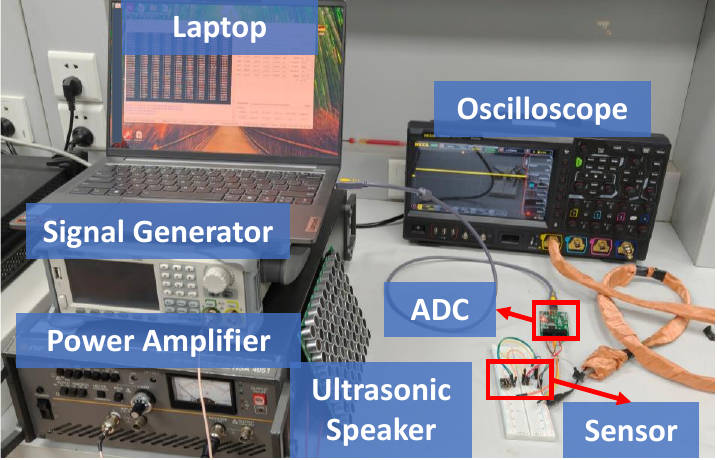}
        \end{minipage}
        \label{fig_setup1}        
    }
    \subfigure[EMI signal emission.]{
        \begin{minipage}[t]{0.45\linewidth}
            \centering
            \includegraphics[width=1\textwidth]{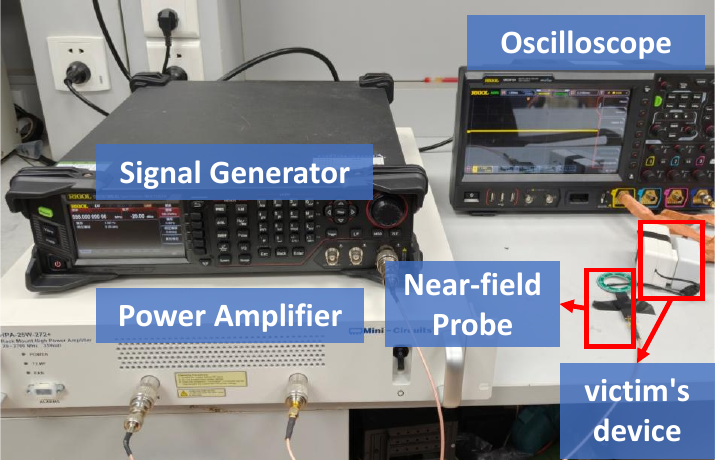}
        \end{minipage}
        \label{fig_setup2}
    }
    \caption{Experimental setup showing attacker devices and victim sensors. The attacker's infrastructure consists of a signal generator, a power amplifier, and signal emitters.}
    \label{fig_setup}
\end{figure}

\begin{figure}[t]
 \centering
 \includegraphics[width=0.97\linewidth]{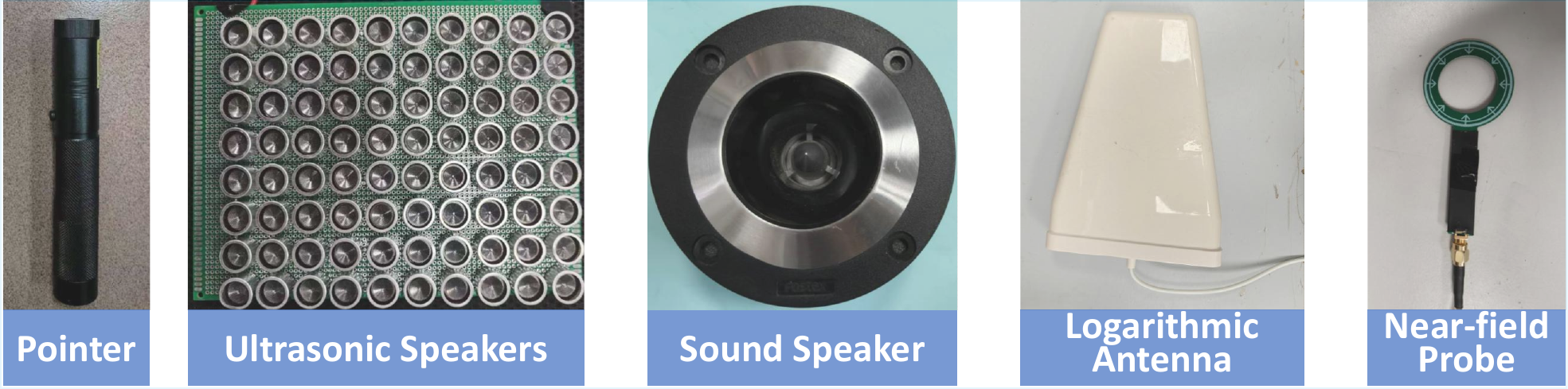}
 \caption{Signal emitters used in our evaluation across different attack modalities.}
 \label{fig_signalemitter}
 \vspace{-5pt}
\end{figure}

\textbf{Attack device configuration.} For optical, ultrasonic, and acoustic attacks, an arbitrary waveform generator~\cite{SIGLENT2021} combined with a power amplifier~\cite{Micronix2013} drives transmitters to emit precise physical signals, using an oscilloscope for real-time monitoring as depicted in \cref{fig_setup1}. Similarly, EMI attacks employ an RF signal generator~\cite{rigol2019} and amplifier~\cite{Mini-Circuits} to drive a logarithmic antenna or near field probe, shown in~\cref{fig_setup2}. 

\textbf{Trigger signal generation.} Addressing real-world deployment, we target remotely deliverable optical, acoustic, and electromagnetic triggers using the emitters displayed in~\cref{fig_signalemitter}.

\begin{figure}[t]
    \centering
    \subfigure[SPV0842LR5H microphone.]{
        \begin{minipage}[t]{0.42\linewidth}
            \centering
            \includegraphics[width=1.0\linewidth]{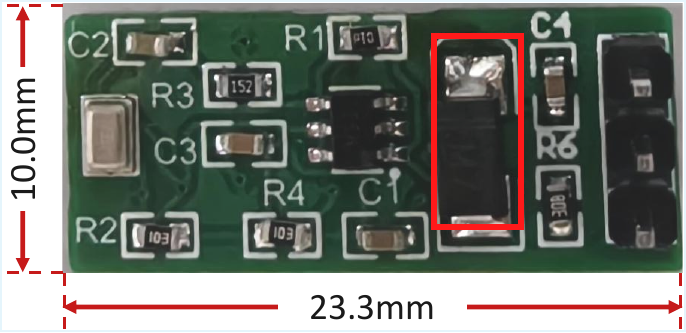}
            
        \end{minipage}
        \label{fig_pcbtrojan1}        
    }
    \subfigure[S150B381 microphone.]{
        \begin{minipage}[t]{0.48\linewidth}
            \centering
            \includegraphics[width=1\textwidth]{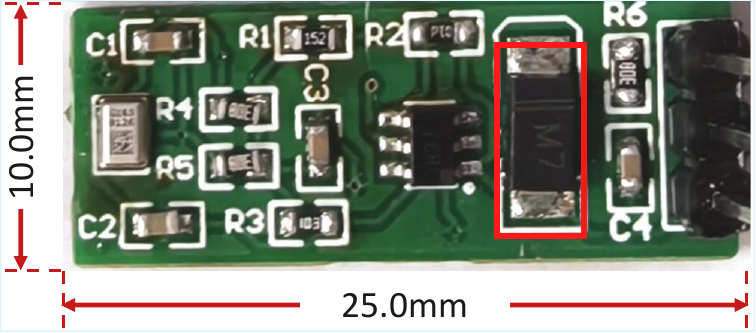}
        \end{minipage}
        \label{fig_pcbtrojan2}
    }
    \caption{PED Trojans embedded in two microphone types using PCB. Red boxes indicate the inserted Trojan circuits.}
    \label{fig_pcbtrojan}
\end{figure}

\textbf{Trojan implementation strategy.} For experimental flexibility, we implement SenTrojan using commercial off-the-shelf sensors and prototype the added Trojan circuitry on circuit boards. These prototypes are intended to validate attack functionality rather than represent the final physical form of a supply-chain implant. In a practical attack, the inserted circuitry can be miniaturized using standard PCB and surface-mount design techniques. To demonstrate feasibility, we further build a PCB-integrated PED Trojan for a microphone, where the complete Trojan-enhanced module measures 10 mm × 25 mm, and the Trojan itself occupies approximately 6 mm × 2 mm, as shown in~\cref{fig_pcbtrojan}. Accordingly, we focus our evaluation on Trojan efficacy and hardware overhead rather than physical miniaturization.


\subsubsection{Evaluation Metrics}
We evaluate the Trojan efficacy of \xxx using three primary metrics and the hardware overhead using four metrics, with detailed metric information as follows:
\begin{itemize}[leftmargin=10pt, nosep]
    \item \textbf{Similarity}  assesses how Trojan affects the sensor’s normal function by comparing the output of the benign sensor with that of the Trojan-inserted one. It is calculated as:
    \begin{equation}
        \text{Similarity} = 1 - \frac{1}{n} \sum_{i=1}^{n} \left| \frac{\Tilde{x}_{i} - x_{i}}{x_{i}} \right|
        \label{eq_similarity}
    \end{equation}
    where $n$ is the total number of the data points, $x_i$ is the value from the benign sensor, $\Tilde{x}_{i}$ is the value from the Trojan-inserted sensor. For complex signals like audio waveforms, we apply frequency-domain transformations and use spectral distance metrics. \rev{In each case, similarity is calculated over all sampled points rather than from a single measurement. For example, the light sensor evaluation collects approximately 108,000 samples over three 10-minute trials under different lighting conditions.}

    \item  \rev{\textbf{False trigger rate (FTR)} is the ratio of benign trials in which the Trojan is unintentionally activated to the total number of benign trials conducted without the predefined trigger signal. For PED cases, a false activation is identified when the output deviation from the benign reference exceeds 5\%, corresponding to the maximum specified measurement error among representative commercial sensors~\cite{NXP_MPX5010, Amphenol_NPC300, Honeywell_MidasS2_H2, Amphenol_A1737, ADI_ADXL321, Infineon_IM72D128V01}.}

    \item \rev{\textbf{Successful trigger rate (STR)} is the ratio of trials in which the Trojan is successfully activated and produces the intended effect to the total number of trials conducted with the predefined trigger signal. We evaluate each Trojan over 50 independent trigger trials and calculate STR.}

\end{itemize}
Since the different effects of LRD Trojans share a common triggering mechanism and differ only slightly in payload circuitry, the evaluation of attack effectiveness and stealthiness is described together. A PED Trojan primarily enables spoofing attacks with minimal circuit modifications.

Details for hardware overhead metrics are as follows:
\begin{itemize}[leftmargin=10pt, nosep]

\item \textbf{Time drift} describes how sensor output changes over time.
\item \textbf{Noise floor} denotes intensity of inherent background noise.
\item \textbf{Response time} shows how fast the sensor reacts to changes.
\item \textbf{Power} represents the electrical power required by the sensor during operation.
\end{itemize}

For hardware overhead, we report the normalized deviation from the benign sensor, where lower values indicate smaller overhead:
    \begin{equation}
    \mathit{Overhead}_m = \frac{1}{n} \sum_{i=1}^{n} \left| \frac{\tilde{m}_i - m_i}{m_i} \right|
        \label{eq_overhead}
    \end{equation}
where $m_i$
 and $\Tilde{m}_{i}$
 denote the benign and Trojan-inserted measurements for metric $m$, respectively.

For each Trojan type, we first introduce the applied sensor and detail the corresponding test procedure used to evaluate Trojan efficacy on a case-by-case basis. We then present the hardware-overhead evaluation in a unified manner.

\begin{table*}[]
\setlength{\tabcolsep}{3pt}
\centering
\caption{Overall performance of \xxx.}
\label{tab:my-table}
\footnotesize
\renewcommand{\arraystretch}{1.3}
\resizebox{\linewidth}{!}{%
\footnotesize
\begin{tabular}{|ccccc|ccc|ccc|cccc|}
\hline
\multicolumn{5}{|c|}{\textbf{\xxx}}                                                                                                                                                                                             & \multicolumn{3}{c|}{\textbf{Sensor}}                                                                   & \multicolumn{3}{c|}{\textbf{Trojan Efficacy}}                                                                             & \multicolumn{4}{c|}{\textbf{Hardware Overhead}}                                                                                    \\ \hline
\multicolumn{1}{|c|}{\textbf{Type}}         & \multicolumn{1}{c|}{\textbf{Cond.}}         & \multicolumn{1}{c|}{\textbf{Mod.}}        & \multicolumn{1}{c|}{\textbf{\makecell{SSI+ \\ DC}}} & \textbf{\makecell{Add.\\ Sensor}} & \multicolumn{1}{c|}{\textbf{Type}}             & \multicolumn{1}{c|}{\textbf{Manuf.}} & \textbf{Model} & \multicolumn{1}{c|}{\textbf{Sim.$\uparrow$}} & \multicolumn{1}{c|}{\textbf{FTR$\downarrow$}} & \textbf{STR$\uparrow$} & \multicolumn{1}{c|}{\textbf{Drift$\downarrow$}} & \multicolumn{1}{c|}{\textbf{Noise$\downarrow$}} & \multicolumn{1}{c|}{\textbf{RT$\downarrow$}} & \textbf{Power$\downarrow$} \\ \hline \hline

\multicolumn{1}{|c|}{\multirow{10}{*}{LRD}} & \multicolumn{1}{c|}{\multirow{5}{*}{Amp.}}  & \multicolumn{1}{c|}{\multirow{5}{*}{Light}}   & \multicolumn{1}{c|}{\multirow{5}{*}{3+8}}         & \multirow{3}{*}{FALSE}          & \multicolumn{1}{c|}{\multirow{3}{*}{Light}}    & \multicolumn{1}{c|}{Adafruit}        & GL5506         & \multicolumn{1}{c|}{99.7\%}                  & \multicolumn{1}{c|}{0\%}                      & 100\%                  & \multicolumn{1}{c|}{0.5\%}         & \multicolumn{1}{c|}{0.3\%}         & \multicolumn{1}{c|}{0.2\%}      & 0.8\%         \\ \cline{7-15} 
\multicolumn{1}{|c|}{}                      & \multicolumn{1}{c|}{}                       & \multicolumn{1}{c|}{}                         & \multicolumn{1}{c|}{}                             &                                 & \multicolumn{1}{c|}{}                          & \multicolumn{1}{c|}{Vishay}          & TEMT6000       & \multicolumn{1}{c|}{99.9\%}                  & \multicolumn{1}{c|}{0\%}                      & 82\%                   & \multicolumn{1}{c|}{0.3\%}         & \multicolumn{1}{c|}{0.5\%}         & \multicolumn{1}{c|}{0.5\%}      & 7.4\%         \\ \cline{7-15} 
\multicolumn{1}{|c|}{}                      & \multicolumn{1}{c|}{}                       & \multicolumn{1}{c|}{}                         & \multicolumn{1}{c|}{}                             &                                 & \multicolumn{1}{c|}{}                          & \multicolumn{1}{c|}{Siemens}         & QRB1           & \multicolumn{1}{c|}{99.4\%}                  & \multicolumn{1}{c|}{0\%}                      & 98\%                   & \multicolumn{1}{c|}{0.5\%}         & \multicolumn{1}{c|}{0.9\%}         & \multicolumn{1}{c|}{0.8\%}      & 6.6\%         \\ \cline{5-15} 
\multicolumn{1}{|c|}{}                      & \multicolumn{1}{c|}{}                       & \multicolumn{1}{c|}{}                         & \multicolumn{1}{c|}{}                             & \multirow{2}{*}{TRUE}           & \multicolumn{1}{c|}{\multirow{2}{*}{Temp.}}    & \multicolumn{1}{c|}{Amphenol}        & NHQ103B        & \multicolumn{1}{c|}{99.0\%}                  & \multicolumn{1}{c|}{0\%}                      & 100\%                  & \multicolumn{1}{c|}{0.9\%}         & \multicolumn{1}{c|}{0.9\%}         & \multicolumn{1}{c|}{4.4\%}      & 13.6\%         \\ \cline{7-15} 
\multicolumn{1}{|c|}{}                      & \multicolumn{1}{c|}{}                       & \multicolumn{1}{c|}{}                         & \multicolumn{1}{c|}{}                             &                                 & \multicolumn{1}{c|}{}                          & \multicolumn{1}{c|}{Heraeus}         & PT100          & \multicolumn{1}{c|}{97.8\%}                  & \multicolumn{1}{c|}{0\%}                      & 96\%                   & \multicolumn{1}{c|}{1.7\%}         & \multicolumn{1}{c|}{0.7\%}         & \multicolumn{1}{c|}{3.4\%}      & 0.5\%         \\ \cline{2-15} 
\multicolumn{1}{|c|}{}                      & \multicolumn{1}{c|}{\multirow{5}{*}{Freq.}} & \multicolumn{1}{c|}{\multirow{5}{*}{Ultras.}} & \multicolumn{1}{c|}{\multirow{5}{*}{3+12}}        & \multirow{2}{*}{FALSE}          & \multicolumn{1}{c|}{\multirow{2}{*}{Mic.}}     & \multicolumn{1}{c|}{CUI Dev.}        & CMC-9745       & \multicolumn{1}{c|}{99.7\%}                  & \multicolumn{1}{c|}{0\%}                      & 96\%                   & \multicolumn{1}{c|}{1.4\%}         & \multicolumn{1}{c|}{0.8\%}         & \multicolumn{1}{c|}{1.4\%}      & 10.2\%         \\ \cline{7-15} 
\multicolumn{1}{|c|}{}                      & \multicolumn{1}{c|}{}                       & \multicolumn{1}{c|}{}                         & \multicolumn{1}{c|}{}                             &                                 & \multicolumn{1}{c|}{}                          & \multicolumn{1}{c|}{DFRobot}         & Sen0487        & \multicolumn{1}{c|}{99.9\%}                  & \multicolumn{1}{c|}{0\%}                      & 98\%                   & \multicolumn{1}{c|}{0.2\%}         & \multicolumn{1}{c|}{0.2\%}         & \multicolumn{1}{c|}{2.4\%}      & 8.7\%         \\ \cline{5-15} 
\multicolumn{1}{|c|}{}                      & \multicolumn{1}{c|}{}                       & \multicolumn{1}{c|}{}                         & \multicolumn{1}{c|}{}                             & \multirow{3}{*}{TRUE}           & \multicolumn{1}{c|}{\multirow{2}{*}{Infrared}} & \multicolumn{1}{c|}{Murata}          & IRA-S210       & \multicolumn{1}{c|}{100\%}                   & \multicolumn{1}{c|}{0\%}                      & 94\%                   & \multicolumn{1}{c|}{0.9\%}         & \multicolumn{1}{c|}{0.7\%}         & \multicolumn{1}{c|}{1.5\%}      & 0.6\%         \\ \cline{7-15} 
\multicolumn{1}{|c|}{}                      & \multicolumn{1}{c|}{}                       & \multicolumn{1}{c|}{}                         & \multicolumn{1}{c|}{}                             &                                 & \multicolumn{1}{c|}{}                          & \multicolumn{1}{c|}{Adafruit}        & SR501          & \multicolumn{1}{c|}{100\%}                   & \multicolumn{1}{c|}{0\%}                      & 96\%                   & \multicolumn{1}{c|}{0.1\%}         & \multicolumn{1}{c|}{0.6\%}         & \multicolumn{1}{c|}{0.9\%}      & 6.7\%         \\ \cline{6-15} 
    \multicolumn{1}{|c|}{}                      & \multicolumn{1}{c|}{}                       & \multicolumn{1}{c|}{}                         & \multicolumn{1}{c|}{}                             &                                 & \multicolumn{1}{c|}{Pressure}                  & \multicolumn{1}{c|}{Dayang}          & DYTB-002       & \multicolumn{1}{c|}{99.3\%}                  & \multicolumn{1}{c|}{0\%}                      & 96\%                   & \multicolumn{1}{c|}{0.3\%}         & \multicolumn{1}{c|}{0.7\%}         & \multicolumn{1}{c|}{0.3\%}      & 3.9\%         \\ \hline
\multicolumn{1}{|c|}{\multirow{4}{*}{PED}}  & \multicolumn{1}{c|}{ED}                     & \multicolumn{1}{c|}{Ultras.}                  & \multicolumn{1}{c|}{0+1}                          & FALSE                           & \multicolumn{1}{c|}{Mic.}                      & \multicolumn{1}{c|}{DFRobot}         & Sen0487        & \multicolumn{1}{c|}{99.8\%}                  & \multicolumn{1}{c|}{\rev{0\%}}                      & \rev{100\%}                   & \multicolumn{1}{c|}{0.2\%}         & \multicolumn{1}{c|}{0.7\%}         & \multicolumn{1}{c|}{0.8\%}      & 7.8\%         \\ \cline{2-15} 
\multicolumn{1}{|c|}{}                      & \multicolumn{1}{c|}{\multirow{2}{*}{Satu.}} & \multicolumn{1}{c|}{\multirow{2}{*}{Sound}}   & \multicolumn{1}{c|}{\multirow{2}{*}{0+1}}         & \multirow{2}{*}{FALSE}          & \multicolumn{1}{c|}{\multirow{2}{*}{Inclin.}}  & \multicolumn{1}{c|}{Analog}          & ADXL335        & \multicolumn{1}{c|}{99.4\%}                 & \multicolumn{1}{c|}{\rev{0\%}}                      & \rev{100\%}                    & \multicolumn{1}{c|}{0.5\%}         & \multicolumn{1}{c|}{0.5\%}         & \multicolumn{1}{c|}{3.5\%}      & 16.5\%         \\ \cline{7-15} 
\multicolumn{1}{|c|}{}                      & \multicolumn{1}{c|}{}                       & \multicolumn{1}{c|}{}                         & \multicolumn{1}{c|}{}                             &                                 & \multicolumn{1}{c|}{}                          & \multicolumn{1}{c|}{Analog}          & ADXL337        & \multicolumn{1}{c|}{99.2\%}                  & \multicolumn{1}{c|}{\rev{0\%}}                      & \rev{98\%}                    & \multicolumn{1}{c|}{0.8\%}         & \multicolumn{1}{c|}{0.9\%}         & \multicolumn{1}{c|}{4.4\%}      & 15.9\%         \\ \cline{2-15} 
\multicolumn{1}{|c|}{}                      & \multicolumn{1}{c|}{Transd.}                   & \multicolumn{1}{c|}{EMI}                      & \multicolumn{1}{c|}{0+0}                          & FALSE                           & \multicolumn{1}{c|}{Temp.}                     & \multicolumn{1}{c|}{People}          & RETCB          & \multicolumn{1}{c|}{99.1\%}                  & \multicolumn{1}{c|}{\rev{0\%}}                      & \rev{96\%}                    & \multicolumn{1}{c|}{1.1\%}         & \multicolumn{1}{c|}{0.5\%}         & \multicolumn{1}{c|}{3.2\%}      & 0.6\%         \\ \hline
\end{tabular}
}

\begin{tablenotes}
    \footnotesize
    \item Mod.: Modality; SSI: Small Scale Integration; DC: Discrete Component; Sim.: Similarity; FTR: False Trigger Rate; STR: Successful Trigger Rate; 
    \item RT: Response Time; Amp.: Amplitude; Freq.: Frequency; ED: Envelope Detection; Satu.: Saturation; Transd.: Transducer; Cond.: Condition; Add.: Additional; Manuf.: Manufacturer; Mic.: Microphone; Temp.: Temperature; Ultras.: Ultrasound; Inclin.: Inclination; CUI Dev.: CUI Devices. 
\end{tablenotes}
\end{table*}

\begin{figure}[t]
    \centering
    \includegraphics[width=1.0\linewidth]{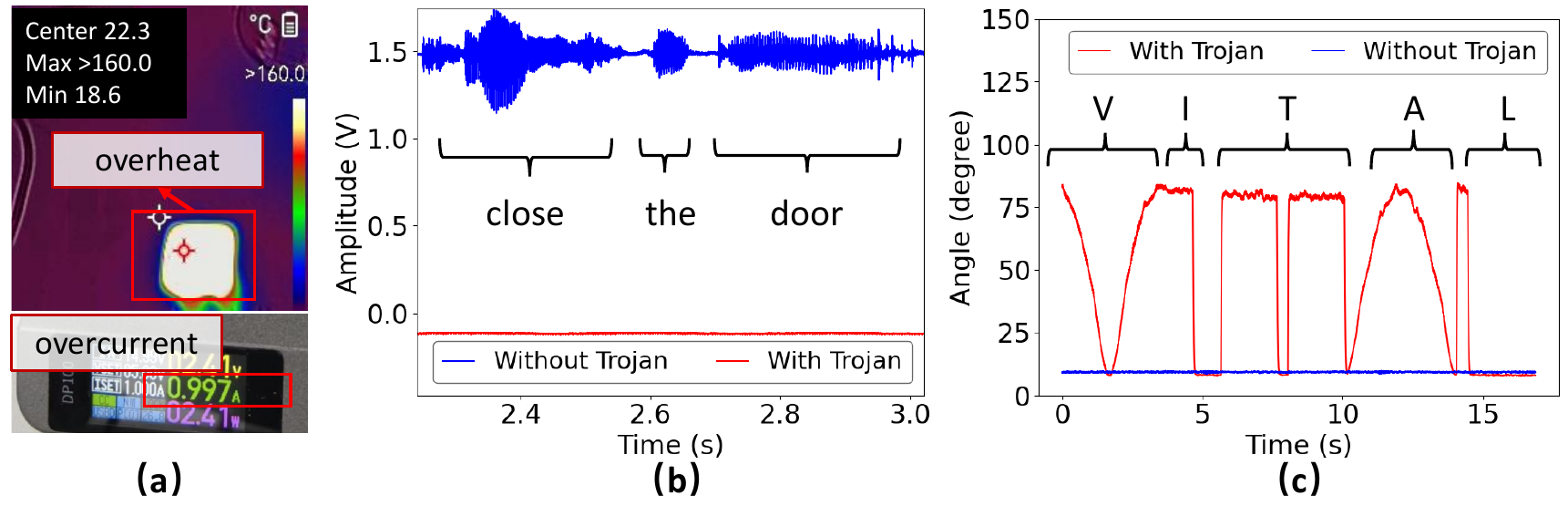}
    \caption{(a) Thermal imaging capture of overheating eventually leading to PDoS, alongside the parameter display of the power supply; (b) Static spoofing results for the microphone; (c) Dynamic spoofing results for the inclination sensor.}
    \label{fig_alleffect}
\end{figure}

\subsection{Amplitude Mechanism of LRD Trojan}
\subsubsection{Ambient Light Sensor}\label{sec_light}
Ambient light sensors are widely used in smart agriculture and industrial combustion systems to measure light intensity. 
 A successful static spoofing attack is defined as the intensity dropping below 10 lux (simulating darkroom or flameout conditions), potentially leading to disrupted crop growth cycles and hazardous fuel accumulation.
We design amplitude detection triggers detailed in Appendix~\ref{sec_app_amp}. 
 We executed 50 trials at 1 m across five diurnal periods, yielding 100\%, 82\%, and 98\% success rates as the laser overcame natural ambient fluctuations. 
 \rev{The STR of TEMT6000 is lower than that of other LRD cases because its transducer package attenuates the injected laser and reduces the effective injection strength. }
 Continuous 72-hour monitoring yielded zero false triggers, while three 10-minute comparative trials under identical lighting revealed average similarities of 99.67\%, 99.88\%, and 99.35\%, confirming negligible operational impact. In addition, the Trojan latch mechanism requires a sensor reset to recover from persistent spoofing.

For PDoS attacks, we employed power supply devices equipped with protective features, including overvoltage protection, overcurrent protection, overheat protection, overload protection, and reverse polarity protection, to ensure testing conditions closely resembled real-world scenarios while avoiding potential risks~\cite{DP100}.
The results are illustrated in~\cref{fig_alleffect}(a). 
\rev{During PDoS attacks, the Trojan-induced short circuit drew 0.99A and raised the local temperature above 170°C. After removing the trigger and power cycling, the sensor did not recover normal functionality.}
Crucially, this thermal breakdown targets the Trojan's own inserted analog switch. Because this engineered short operates outside the sensor's native current path, localized overcurrent protections are inherently bypassed. Unless strict global current limiters are enforced at the main power supply, native safeguards fail to prevent this localized destruction. Furthermore, even if modern industrial devices are equipped with robust current-limiting regulators and thermal protections, the Trojan effectively exploits these very safeguards. By intentionally driving the device into a sustained state of thermal or overcurrent stress to trigger continuous protection cycles, the Trojan achieves a persistent DoS attack, successfully disabling the sensor even when irreversible physical destruction is prevented.

\subsubsection{Temperature Sensor}
In addition, we employ this Trojan mechanism on a temperature sensor, incorporating a photodiode as an extra transducer. This Trojan executes static spoofing by clamping the sensor reading to a falsified low temperature, thereby forcing the thermal control system into a continuous heating state, potentially causing overheating. The trigger success rate was assessed following the identical methodology used for the light sensor.
For stealthiness evaluation, continuous sampling was conducted from 20~°C to 80~°C on a controlled heating platform. The results showed no false triggers. Specifically, for NHQ103B and PT100, the Trojan-inserted versions exhibited data similarities of 98.95\% and 97.79\%, respectively, compared to their benign baselines.


\subsection{Frequency Mechanism of LRD Trojan}

\subsubsection{Microphone} \label{sec-micro1}
Leveraging rare high-intensity ultrasound during normal microphone operation, we design frequency detection triggers detailed in Appendix~\ref{sec_app_freq}. \xxx can silence microphone output to 0~V, as shown in~\cref{fig_alleffect}(b), yielding 96\% and 98\% success rates across 50 trials at a 20 cm distance for two different sensor models. 
Maximum attack distance evaluations using identical ultrasound injections reveal that the Trojan-inserted Sen0487 achieved a 5.1 m DoS range versus 0.5 m for standalone attacks at equal power, whereas the CMC-9745 was immune to DoS attacks without the Trojan.
Stealthiness tests simulating household environments with MUSAN audio~\cite{snyder2015musan} of 10 hours produced no false triggers. Similarity between benign and Trojan-inserted sensors for Sen0487 and CMC-9745 exhibited 99.9\% and 99.7\%, respectively, across 2 hours of recording via normalized log spectral distance~\cite{gray1976distance}, confirming minimal operational impact.





\subsubsection{Trojan Application}
Integrating an extra microphone transducer enables ultrasonic triggering for other sensors.

\begin{figure}[t]
 \centering
 \includegraphics[width=1.0\linewidth]{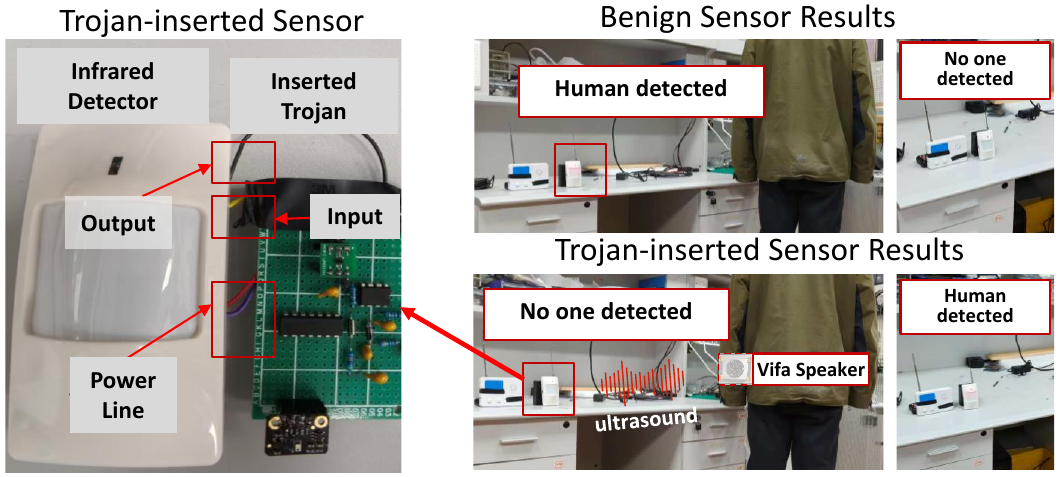}
 \caption{Infrared intrusion-detection experiment. The red indicator denotes human-presence detection; after Trojan activation, the detector reports an incorrect state.}
 \label{fig_infare_realworld}
\end{figure}

\textbf{Infrared sensor.}\label{sec_infrared} Infrared sensors detect human presence via a digital output generated from an infrared-sensitive resistor and a comparator.  The infrared sensor is installed at an office doorway and monitors over a continuous 4-hour period to evaluate the false trigger rate and similarity.  The sensor accurately identified all 47 individuals with zero false positives, maintaining 100\% average similarity to the benign baseline under normal operation.
In addition, we employ this Trojan in a commercial intrusion detection system that uses infrared sensors to detect intruders and trigger alarms. 
At a distance of 1 meter, the attacker used a Vifa speaker~\cite{Vifa} to emit the ultrasonic trigger signal.
Under this setup, the Trojan-inserted sensor could be induced to either miss a human-presence event and output “No one detected,” or falsely report a human presence, whereas the system reported correct detection under normal operation, as shown in~\cref{fig_infare_realworld}. These effects enable attackers to evade infrared detection and potentially conduct unauthorized activities such as trespassing or accessing restricted areas.

\begin{figure}[t]
 \centering
 \includegraphics[width=1.0\linewidth]{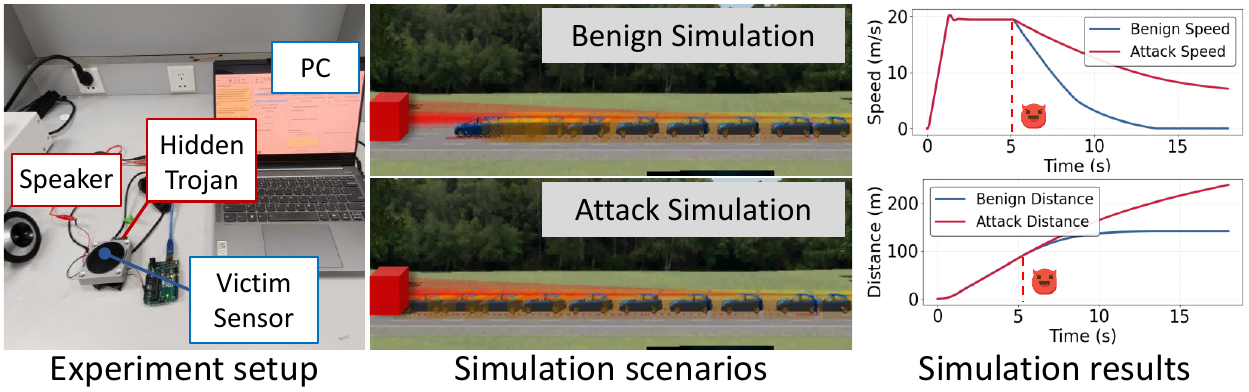}
 \caption{Evaluation of a pressure brake sensor with a HIL driving simulation.}
 \label{fig_weight}
\end{figure}

\textbf{Pressure sensor.}
Linearly indicating brake magnitude, the pedal pressure sensor constitutes a critical component for safe driving. We develop a Hardware-In-the-Loop (HIL) simulation utilizing CarSim~\cite{MechanicalSimulationCorporationCarSim2023} and MATLAB Simulink~\cite{MathWorksHardwareInTheLoop2023} to illustrate the control-level consequence of the Trojan in a closed-loop simulation environment.
In this setup, the pressure sensor output is sampled by an Arduino Uno and forwarded to the PC-based vehicle controller, allowing us to compare benign and Trojan-triggered conditions under the same control logic and obstacle scenario.
For the benign condition, the vehicle initially travels at a constant speed, and braking is applied when an obstacle is detected ahead. With the benign sensor input, the controller receives a valid brake signal, the vehicle speed decreases to zero as expected, and the car stops before reaching the obstacle.
For the attack condition, once the Trojan is triggered, the compromised pressure sensor output is forced to 0, causing the controller to miss the braking event. Under this controlled HIL configuration, the simulated vehicle therefore fails to decelerate in time and collides with the obstacle. The experimental setup, along with the pre-collision speed and travel distance profiles, is shown in~\cref{fig_weight}.
  

\subsection{The PED Trojan of Microphone}\label{sec_micro2}
\begin{figure}[t]
   \centering
   \begin{minipage}[b]{0.42\linewidth}
      \includegraphics[width=1.\linewidth]{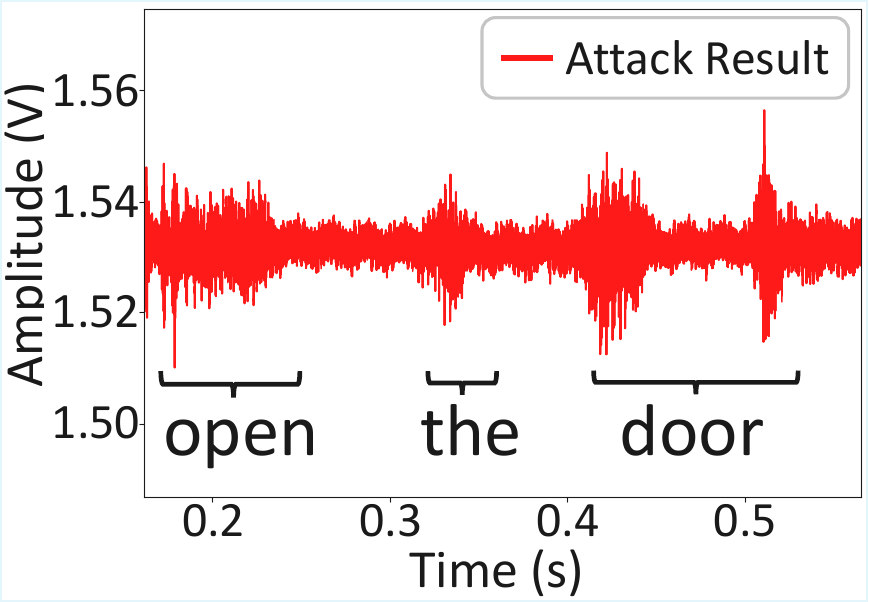}
      \caption{ The waveform of the injected command to spoof the microphone.}
      \label{fig-command}
      
   \end{minipage}\qquad
   \begin{minipage}[b]{0.45\linewidth}
      \includegraphics[width=1.\linewidth]{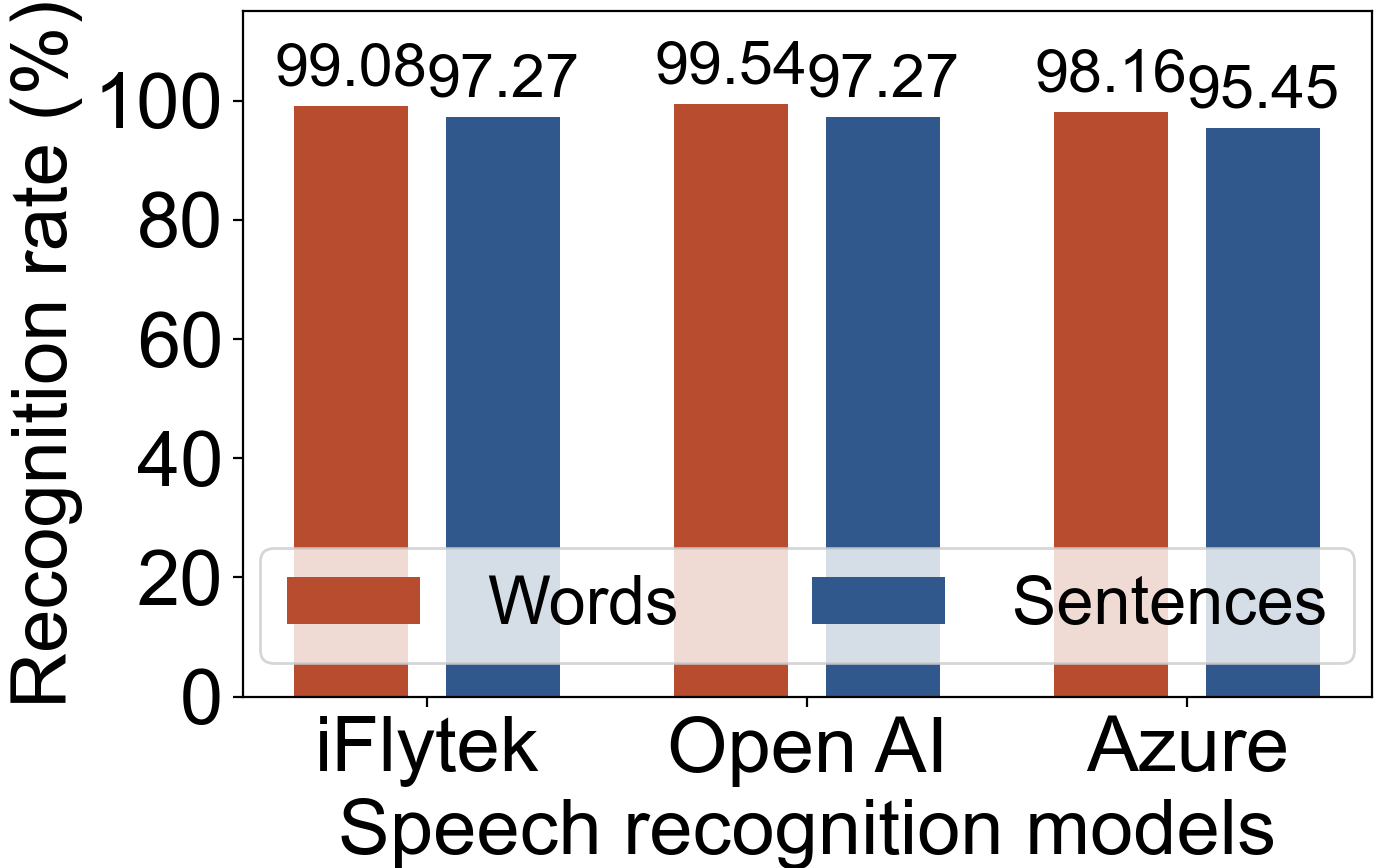}
      \caption{The recognition result of the Trojan-based attack on three ASR models.}
      \label{fig-asr}
   \end{minipage}
\end{figure}

PED Trojans utilize amplitude-modulated voice embedded in carriers to spoof sensors. During Trojan activation, they demodulate high-frequency ultrasound to recover baseband commands like opening the door, potentially triggering unauthorized actions as shown in~\cref{fig-command}. Our setup employs computer-generated voice basebands and 27 kHz carriers to evaluate effectiveness following NormDetect protocols~\cite{li2023learning,yang2025light}. Testing recognition across IFlytek~\cite{iflytek}, OpenAI~\cite{openai}, and Azure~\cite{Microsoft} ASR systems yielded results in~\cref{fig-asr}. To mitigate nonlinearity, we first injected 27 kHz electrical signals directly into amplifier ports, achieving 100\% success. Second, leveraging prior ultrasonic methods~\cite{zhang2017dolphinattack}, we reduced sampling rates to 16 kHz, sufficient for speech. As ultrasonic components alias through the ADC to eliminate nonlinear impacts, we distinguished Trojan-induced behavior from system functionality, demonstrating 100\% success. Conversely, benign microphones yielded only noise under identical injection due to weak demodulation. \rev{Finally, we assessed similarity and STR using the same PED Trojan evaluation protocol. The microphone achieved a 100\% STR over 50 trigger trials. By placing the device separately in laboratory and office environments for ten hours, we observed no false ASR sentence activations (FTR = 0\%).}

\textbf{Application.} 
Modern cars are equipped with voice virtual assistants that use ASR models to interpret user commands. To demonstrate the threat of microphone spoofing, we conduct an experiment where commands demodulated by the Trojan are injected into the victim's microphone in the car to closely simulate real attack scenarios. The results reveal that the injected command causes the car door to open, potentially leading to dangerous situations, such as unauthorized access to the vehicle or theft.


      

\subsection{The PED Trojan of Inclination Sensor}  \label{sec_inclination}
Extensively deployed in drones, smartphones, and automotive safety systems~\cite{fisher2010using}, inclination sensors measure acceleration and angles.
We investigate PED Trojan designs targeting modules dependent on signal processing circuits.
Frequency sweeps identify z-axis resonance at 720 Hz for the ADXL335 and 518 Hz for the ADXL337. Controlled by trigger intensity and amplifier asymmetry, DC amplitude manipulation is shown in~\cref{fig_alleffect}(c). \rev{\xxx induces 159° and 26° biases for the tested sensor models, which are the maximum sensor-level deviations measured in our PED Trojan experiments.} 
\rev{Because PED Trojans lack the well-defined activation threshold characteristic of LRD Trojans, we first establish a reference output in a preliminary test. We then define that the PED Trojan is successfully activated if the result of a subsequent test falls within $\pm 20\%$ of this reference value.}
Conversely, benign sensors exposed to identical triggers exhibit no distortion due to insufficient asymmetry for saturation. Finally, comparing benign and Trojan-inserted sensors across nine angles yielded 99.43\% and 99.23\% similarity, confirming low observability under normal operation. 
\rev{We further evaluated the FTR across nine angles, repeating each measurement five times and collecting 10~s of data at each angle. The results showed output similarities of 99.4\% and 99.2\% for the ADXL335 and ADXL337, respectively, corresponding to deviations below 5\%, with an FTR of zero for both sensors. Over 50 trigger trials, the ADXL335 and ADXL337 achieved STRs of 100\% and 98\%, respectively.}

\begin{figure}[t]
 \centering
 \includegraphics[width=1.0\linewidth]{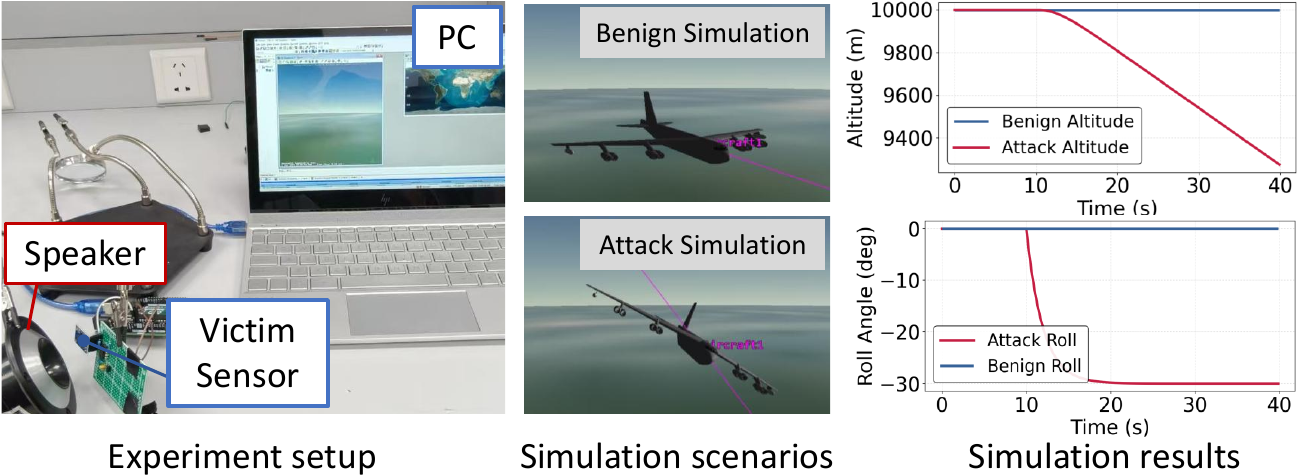}
 \caption{Evaluation of the inclination sensor in the flight control system. When the Trojan is triggered, the aircraft undergoes an attitude tilt and its altitude decreases rapidly.}
 \label{fig_angle}
\end{figure}

\textbf{Application.}
Widely applied in robotics and aviation, inclination sensors can directly affect flight stability and safety, and constitute potential factors in crashes~\cite{curt2019boeing}.
We compute the position coordinates and body-attitude quaternions from the sensor outputs and import them into STK~\cite{AnalyticalGraphicsIncSTK2025} for flight simulation under matched benign and attack conditions. 
\rev{We feed ADXL335 measurements into the STK simulation and introduce a Trojan-induced 30° attitude deviation, which is sufficient to disrupt attitude control and initiate a right roll.}
Under the benign condition, the aircraft maintains a stable attitude and altitude. Under the attack condition, the erroneous sensor measurement causes the flight controller to issue opposing corrective commands, producing a right roll. We further incorporate a lift-loss model to capture the effect of the increasing roll angle on altitude. As shown in~\cref{fig_angle}, the attack induces an attitude tilt and rapid altitude loss.



\subsection{The PED Trojan of Temperature Sensor} \label{sec_temperature}
Temperature sensors are critical in medical and industrial systems, where erroneous readings can disrupt closed-loop thermal control and contribute to unsafe operating conditions, potentially including excess cooling scenarios such as the British dye factory incident~\cite{process2018excess}.
Targeting temperature-controlled outlets that regulate power based on real-time thermal feedback, we examine a PED Trojan mechanism implemented through transducer wire-length modification, which is common in chemical reactors and industrial manufacturing~\cite{InstrumartTempControllerBasics2018, WestControlSolutionsTemperatureMonitors2018}.

\begin{figure}[t]
    \centering
    \subfigure[Setup.]{
        \begin{minipage}[t]{0.25\linewidth}
            \centering
            \includegraphics[width=1.0\linewidth]{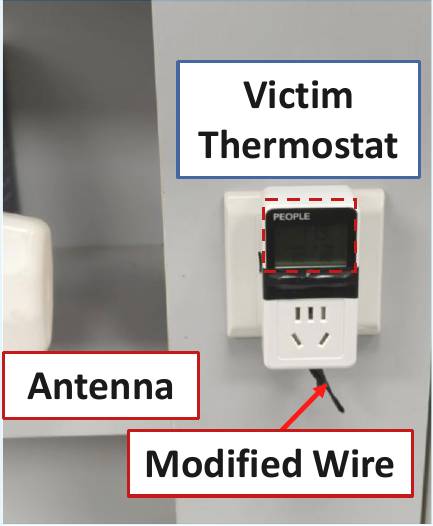}
        \end{minipage}
    }
    \subfigure[Result.]{
        \begin{minipage}[t]{0.65\linewidth}
            \centering
            \includegraphics[width=1\textwidth]{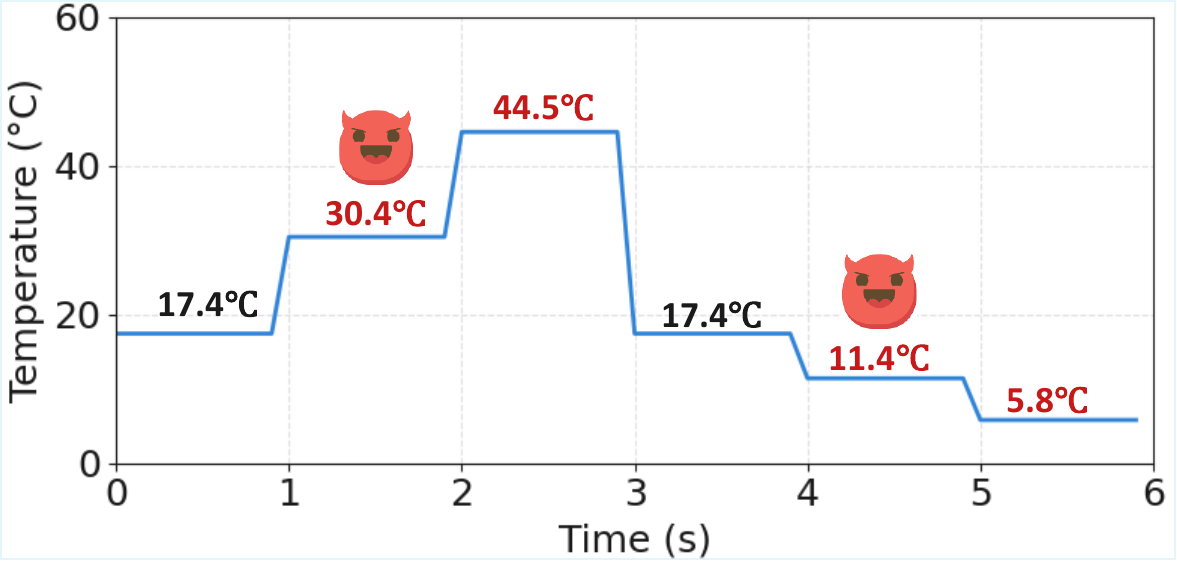}
        \end{minipage}
    }
    \caption{The temperature-controlled system with wire length adjustment serving as the Trojan mechanism.}
    \label{fig_temperature}
\end{figure}

Modifying transducer wire lengths enables frequency-dependent attacks. Specifically, 10 cm wires yield 11.6°C drops at 789 MHz and 27.1°C increases at 710 MHz, while 1 cm wires exhibit 2.2~°C drops and 5.6~°C increases at 1185 MHz and 1298 MHz, respectively, as illustrated in \cref{fig_temperature}. This enhancement amplifies the variation by 4.9× without hardware component changes. Furthermore, adjusting signal amplitude controls deviations, enabling attackers to trigger thermal runaway, production hazards, or financial loss. 
\rev{We further evaluated the FTR across ten temperatures, repeating each measurement five times. The results showed an output similarity of 99.1\%, corresponding to a deviation below 5\%, with an FTR of zero. Over 50 trigger trials, the RETCB temperature sensor achieved an STR of 96\%.}

\subsection{Hardware Overhead}
To evaluate the hardware overhead of \xxx, we compare the factory-calibrated performance of benign and Trojan-inserted sensors under identical environmental and physical stimuli. For each metric, we report the normalized deviation measured under the same test condition. The resulting deviations are summarized in~\cref{tab:my-table}, where lower values indicate smaller functional perturbation and better stealth.

\textbf{Time drift.} 
We collect 600 samples at 10 Hz immediately after initialization and another 600 samples after 24 hours of continuous operation, and compare the corresponding sensor outputs under identical conditions.
The comparison shows no long-term DC offset beyond the baseline variation, with the measured normalized deviation remaining below 2\% across all tested sensors. This result is consistent with the high-impedance design of the LRD monitoring circuit, which reduces loading on the native signal path. The added parasitic capacitance is also small relative to the circuit capacitance, limiting its impact on long-term output stability.

\textbf{Noise floor.} 
We quantify the noise floor using the average peak-to-peak voltage measured under identical test conditions. The measured noise-floor deviation remains below 1\% across all tested sensors. For LRD Trojans, the comparator inputs operate in a high-impedance state, which limits leakage-induced current noise. For PED Trojans, the low-frequency power-supply rejection of the analog front end attenuates noise introduced by the added series components, limiting its effect within the measured operating bandwidth.

\textbf{Response time.} 
We evaluate the response time using the 10\% to 90\% step-response rise time. The measured normalized deviation ranges from 0.2\% to 4.4\% across the tested sensors. For LRD Trojans, the added series analog switch contributes only $R_{on} \approx 6\ \Omega$, and the parallel monitoring circuits do not modify the original sensor signal path, resulting in little change to the output $RC$ time constant. 
For PED Trojans, the inserted series resistors and diodes have limited impact on transient response. The resistors keep transient currents within the operational amplifier's driving capability, the diodes add little dynamic resistance during current spikes, and the analog front end compensates for small supply-voltage drops.
These design choices reduce Trojan-induced circuit overhead and match the measured rise-time results.

\textbf{Power.} 
We evaluate power overhead using a DC power supply to monitor sensor power consumption. For LRD Trojans, the normalized power overhead remains below 10\% on average. This result is consistent with the use of LinCMOS components and mega-ohm resistors, which reduce power consumption and keep the Trojan power consumption at the microwatt level. For PED Trojans, the power change depends on the insertion mechanism. Inserted diodes can slightly reduce overall power consumption by lowering the downstream load voltage, whereas some series-resistor PEDs produce a normalized power deviation above 10\% in low-power sensors. Nevertheless, sensor power consumption can vary substantially under different physical stimuli during normal operation because different stimulus intensities induce different current levels. For example, a microphone can consume more than twice the power measured under no-sound input, showing that the observed PED overhead is comparable to normal stimulus-dependent power variation.

%% file: sections/discussion.tex
\section{Potential Defenses}
We propose passive shielding and active detection to defend against physical signal injection in \xxx.

\subsection{Passive Shielding}

\begin{figure}[t]
 \centering
 \includegraphics[width=1.0\linewidth]{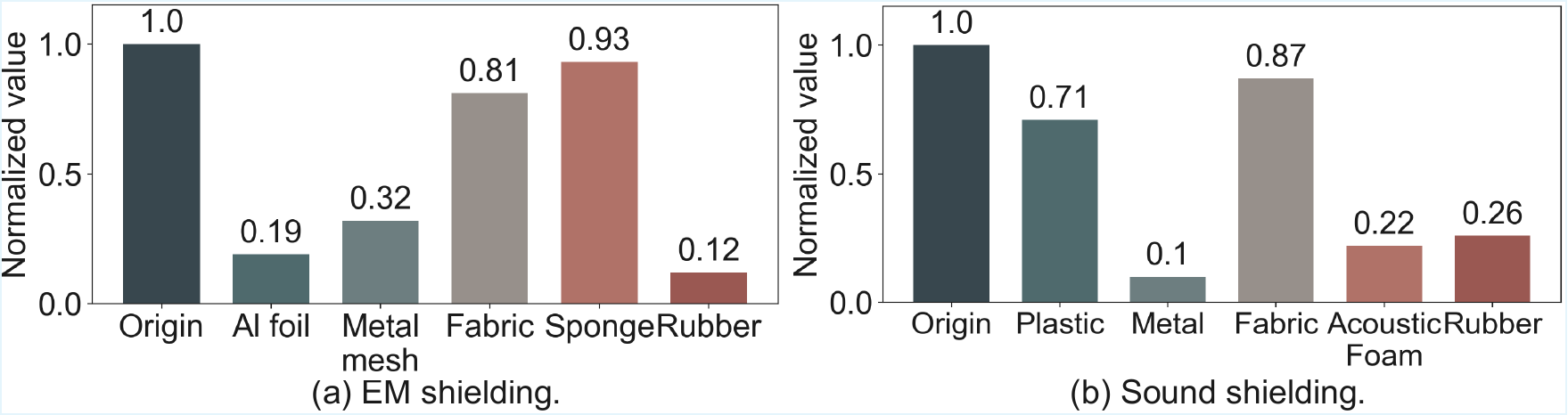}
 \caption{The shielding effects of different materials.}
 \label{fig_shield}
\end{figure}

As the primary defense against EM, acoustic, and optical signals, passive shielding uses metal casings to block EM coupling, opaque materials to prevent lasers, and sound-absorbing foam to dampen acoustic waves, as shown in~\cref{fig_shield}.


\subsection{Active Detection}
Active protection employs EM leakage analysis and functional testing to distinguish benign from Trojan circuits, utilizing runtime anomaly detection to identify triggers.

\textbf{Active detection based on EM leakage.} Detection proves difficult as our attack exploits (1) triggers activating signals beyond normal ranges, (2) high-impedance parallel or short-circuit series inactivity modes minimizing power and circuit performance impact, and (3) subtle parametric modifications obscured by component density. EM-based detection methods exploit circuit-level modifications introduced by sensor Trojans, which alter the resulting electromagnetic leakage profiles. Clustering algorithms then use these leakage patterns to distinguish benign circuits from Trojan-infected ones. However, this approach is ineffective against PED Trojans because their small component-parameter perturbations produce negligible changes in EM leakage.


\textbf{Detection based on functional testing.} Extending standard evaluations~\cite{mcfadden2010supply}, functional testing exposes physical triggers by varying amplitudes, frequencies, and modalities, including acoustic, optical, and electromagnetic. Conversely, duration- or pattern-dependent activations evade detection due to impractical requirements for long-duration injections.

\begin{revision}
    
\textbf{Impedance-based tamper detection}.
For pre-deployment verification, an LRD Trojan employs high-impedance parallel monitoring. Its added components may alter the overall circuit impedance, enabling detection using methods~\cite{mosavirik2022scatterverif} when a trusted reference profile is available.
A PED Trojan minimizes component modifications and therefore causes smaller impedance perturbations. For example, a modification that changes only the wire length may be difficult to detect.
For runtime impedance-based monitoring, an activated LRD Trojan toggles the signal path, thereby changing the circuit impedance. This change can potentially expose the Trojan to impedance monitoring against a trusted reference.
\end{revision}

\textbf{Runtime anomaly detection.} 
Complementing pre-application detection, runtime verification targets correlation errors and spectral noise anomalies, mitigating prohibitive physical isolation costs. Utilizing sensor fusion for consistency checks while scanning transducer outputs for distortion from common injections, including out-of-band frequencies, this monitor effectively identifies attack footprints.

%% file: sections/related_work.tex
\section{Related Work} \label{related_work}

\subsection{Sensor Attacks}

Prior research has revealed that a wide variety of sensor types across multiple fields are vulnerable to security threats.  For example, modulated lasers can inject voice commands into microphones~\cite{sugawara2020light}, while electromagnetic interference can induce false touch events on capacitive touchscreens~\cite{jiang2022wight}. These threats have been demonstrated in automotive systems~\cite{shin2017illusion,petit2015remote,chauhan2014platform,xu2018analyzing}, drones~\cite{son2015rocking,trippel2017walnut}, smart devices~\cite{zhang2017dolphinattack}, energy systems~\cite{selvaraj2018electromagnetic}, medical devices~\cite{kune2013ghost}, and other applications~\cite{guri2017led}. This research often involves introducing attack signals through physical media, such as light~\cite{sugawara2020light}, acoustic waves~\cite{zhang2017dolphinattack}, electromagnetic waves~\cite{kune2013ghost}, or electric fields~\cite{maruyama2019tap}. 
Despite extensive research uncovering diverse sensor attack vectors, existing approaches remain fundamentally constrained by their reliance on pre-existing hardware vulnerabilities and continuous external signal emission. As a result, their effectiveness is often sensor-specific, and their impact is difficult to sustain once the attack signal is removed. In contrast, \xxx investigates internal hardware compromises via sensor Trojans, which embed malicious functionality into sensor circuits to make previously resistant sensors susceptible to crafted physical triggers. This shift from exploiting to implanting vulnerabilities broadens the sensor attack surface, enabling more generalizable, controllable, and persistent attacks.




\subsection{Hardware Trojan Attacks}
Hardware Trojans, typically realized through malicious circuit modifications~\cite{tehranipoor2010survey}, have been widely studied in IC security, particularly for high-stakes military systems. Prior work examined untrusted military ICs~\cite{collins2006trust} and kill-switch threats~\cite{adee2008hunt}, both involving digital circuitry insertion. Compared with IC-level Trojans, board-level security remains less explored~\cite{zhu2021pcbench}, motivating automated synthetic Trojan insertion into PCB designs~\cite{bhattacharyay2022automated}. Existing Trojans are commonly classified as logic-based designs~\cite{xiao2016hardware,bhunia2014hardware}, which add gates, or dopant-based designs~\cite{becker2013stealthy,hu2020overview}, which modify physical properties.
While Yang et al.~\cite{yang2016a2} replace digital counters with analog capacitors for Trojan triggering, their design still monitors internal digital states and targets digital logic. In contrast, \xxx targets analog sensors, where crafted external physical signals activate hidden analog behavior without software or digital access. \rev{Comparisons of \xxx with A2~\cite{yang2016a2}  are shown in Appendix~\ref{sec_app_A2}.}  This shift introduces analog-specific challenges, namely preserving benign sensing accuracy while ensuring reliable activation under environmental noise and component variations. Moreover, by leveraging native physical effects such as transducer nonlinearity and amplifier saturation, \xxx reduces reliance on explicit logic anomalies and can evade conventional digital-integrity checks and visual inspections.

%% file: sections/conclusion.tex
\section{Conclusion}
In this paper, we introduce \xxx, a novel threat vector manipulating sensor outputs via malicious hardware modifications. We propose two distinct trigger mechanisms: Logic-Rule-Driven (LRD) Trojans relying on dedicated blocks for specific signal patterns, and Physical-Effect-Driven (PED) Trojans exploiting inherent behaviors. Extensive evaluations across 14 sensors confirm reliable triggering with minimal operational disruption. Finally, we explore countermeasures mitigating impacts, emphasizing designer vigilance regarding this emerging threat.


%% file: sections/appendix.tex
\section{Appendix}

\subsection{Life Cycle of Sensor} \label{app-lifecycle}
Sensor life cycles typically span five phases: design, fabrication, distribution, maintenance, and recycling~\cite{bhunia2014hardware,tartagni2022electronic,webster2018measurement,xiao2025sok}. (1) Design: Engineers define structures, component parameters, and types for chips and peripheral circuits based on application requirements. (2) Manufacturing: Following design finalization, factories produce chips and board-level circuitry. While final testing ensures standards, cost constraints often limit manufacturers to basic functionality checks, omitting thorough safety evaluations~\cite{mcfadden2010supply}. (3) Distribution: Supply chains deliver standalone chips or fully assembled modules for system integration. (4) Maintenance: High-end sensors require regular calibration, diagnostics, and component replacement to ensure consistent long-term performance. (5) Recycling: End-of-life recovery and refurbishment support cost savings and environmental sustainability.

\subsection{The Adversary's Capabilities of Different Stages} \label{app_insert}

\begin{itemize}[leftmargin=10pt, nosep]
    \item \textbf{Design stage.} One attack scenario arises when malicious actors infiltrate the sensor design team and obtain white-box access to circuit parameters and component choices, a threat model widely explored in prior work~\cite{zhang2014detrust,waksman2011silencing,trippel2021bomberman,imeson2016non}. Specifically, malicious insiders covertly modify hardware components to embed Trojans before delivery. This stage is feasible because sensor design, verification, and subsequent fabrication involve different personnel or organizations. However, the modification must evade design review, calibration, and functional validation, while preserving sensing range, noise, drift, power, and normal behavior. Routine black-box tests may not exercise a hidden trigger, but an independent golden model may reveal unauthorized changes. However, the golden model is unavailable when the same party controls both the modification and the reference artifacts.
\item \textbf{Manufacturing stage.}
Another attack scenario occurs during sensor fabrication~\cite{yang2016a2,dharsee2023jinn,de2020hartbleed}, when an untrusted foundry manufactures the sensing chip. Adversaries with access to fabrication masks or process parameters can introduce layout- or process-level modifications as Trojans without changing the overall circuit architecture. This stage is feasible because fabrication is often separated from sensor design and may be performed by an untrusted party. However, analog modifications must remain effective despite process variation while preserving reliability and normal sensing behavior. Moreover, a trusted design file may enable physical or electrical comparison and expose discrepancies.
\item \textbf{Post-manufacturing stage.} 
Attackers may also insert Trojans after manufacturing is complete~\cite{shwartz2017shattered,mitre_attack,ico_supply_chain_2023}, including during PCB assembly and maintenance. In modern sensor systems, discrete sensing elements, signal-conditioning circuits, and sometimes microcontrollers are integrated on the PCB. Attackers involved in assembly or maintenance may replace legitimate components with compromised ones, insert additional circuitry, or modify PCB traces to manipulate sensor outputs. This stage is feasible because assembly, repair, calibration, and maintenance provide direct access to board-level components and interconnects. However, the attacker must overcome physical-access constraints, enclosure restrictions, limited board area, and the need to preserve normal operation after rework. Because the sensor has typically completed factory-side hardware inspection and validation by this stage, a Trojan that preserves normal functionality and avoids unintended activation is difficult to identify through routine post-service testing. 
\end{itemize}  

\subsection{Trigger Signal Modality} \label{app_triggersignal} 
We evaluate ultrasound, laser, EMI, and sound signals across five critical dimensions: flexibility, distance, effect, generality, and difficulty.

\begin{itemize} [leftmargin=10pt, nosep] 
\item \textbf{Trojan flexibility.} Signal modality dictates injection nature and design constraints. Laser offers superior flexibility, injecting fixed-intensity and modulated signals. Ultrasound and sound rank secondary, limited to modulated signals, while EMI exhibits the lowest flexibility, as high-frequency signals induce capacitive and inductive failures.

\item \textbf{Attack distance.} Laser achieves maximum range via minimal attenuation, followed by directional ultrasound. Conversely, sound propagates omnidirectionally, while EMI relies on restricted near-field coupling.

\item \textbf{Attack effect.} Trigger modalities influence outcomes. Laser, ultrasound, and sound support broad attack spectra, including DoS and electrical faults. Conversely, EMI is limited to spoofing as high-frequency behaviors alter component responses, reducing versatility.

\item \textbf{Attack generality.} EMI demonstrates the highest generality as most sensors possess coupling accessible transmission lines. In contrast, other modalities rely on specific transducer types, limiting injection efficiency and applicability.

\item \textbf{Attack difficulty.} Laser attacks present the greatest challenges due to visibility and targeting precision. While audible and detectable, sound simplifies deployment via omnidirectional propagation. Finally, ultrasound and EMI offer maximum stealth, remaining inaudible with relaxed targeting requirements. 
\end{itemize}

\subsection{Analysis of Saturation} \label{appen-satu}

The output range of an amplifier is determined by the value of the power supply and the circuit structure, where the maximum output value is $V_{max}$, and the minimum value is $V_{min}$.   Saturation is common in analog circuits and refers to a quantity that cannot exceed either $V_{max}$ or $V_{min}$.   
The quiescent point is the DC bias $V_q$ required for an amplifier to make the transistor operate in the linear region. It is necessary for the proper operation of the amplifier.
There is no fixed standard value for the bias voltage in single-supply amplifiers, which depends on the specific type and application scenario. The voltage corresponding to the DC quiescent point is usually set to $({V_{max}+V_{min}})/2$ to provide the maximum dynamic range.
The output signal of the amplifier is:
    \begin{equation} \label{euqation_amplifier}
            V_{aout} = 
            \begin{cases}
            V_{min} & \text{if }  A_{d}V_{in}+V_{q} \leq V_{min}\\
            A_{d}V_{in}+V_{q}  &  \text{if } V_{min} \leq A_{d}V_{in}+V_{q} \leq V_{max}\\
            V_{max} & \text{if }  A_{d}V_{in}+V_{q} \ge V_{max}
            \end{cases} 
    \end{equation}
where $A_d$ is the differential gain of the input voltage and $V_{in}$ is the amplitude of  input signal. 

When the injected AC signal exceeds the dynamic range, the output signal will typically be symmetrically clipped( represented as $V_{max}-V_{q}=V_{q}-V_{min}$) by the amplifier due to the design of the quiescent point. 
The AC signal will then be removed by the low-pass filter (LPF).

\subsection{The Circuit Simulation Details}
To better understand the function of different components intuitively, we simulate the circuits using the PSpice model~\cite{pspice}.

\subsubsection{Frequency Detector Simulation}
\begin{figure}[H]
    \centering
    \includegraphics[width=1.0\linewidth]{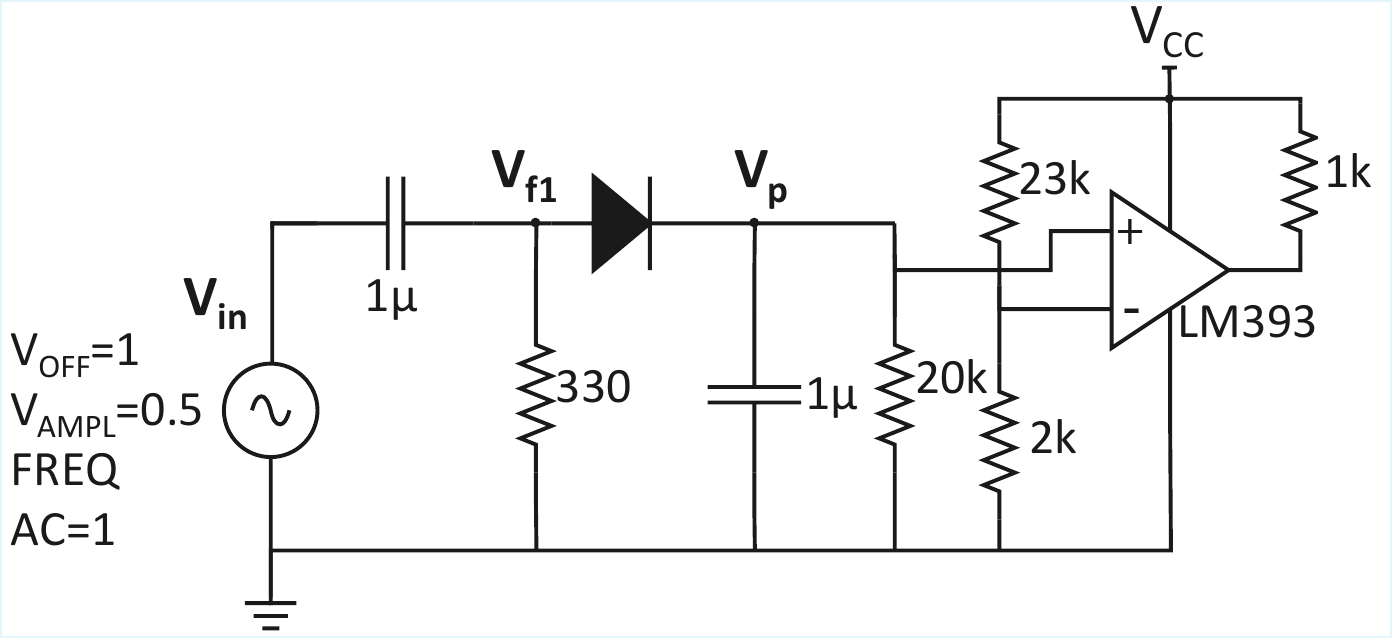}
    \caption{ The PSpice simulation circuit of the frequency detector. }
    \label{fig_frequency_circuit}
    \vspace{-5pt}
\end{figure}

We simulate the trigger mechanism using an LM393 comparator as illustrated in~\cref{fig_frequency_circuit}. The circuit employs a front-end rectifier to convert the AC input into a DC voltage Vp. When the input amplitude generates a Vp exceeding the reference threshold set by the resistor network, the comparator output toggles. This mechanism confirms that injected AC signals can effectively manipulate the digital output state without affecting normal operation at lower amplitudes.

\subsubsection{The Saturation Effect Simulation} \label{sec_sim_amp1}

\begin{figure}[H]
    \centering
    \includegraphics[width=1.0\linewidth]{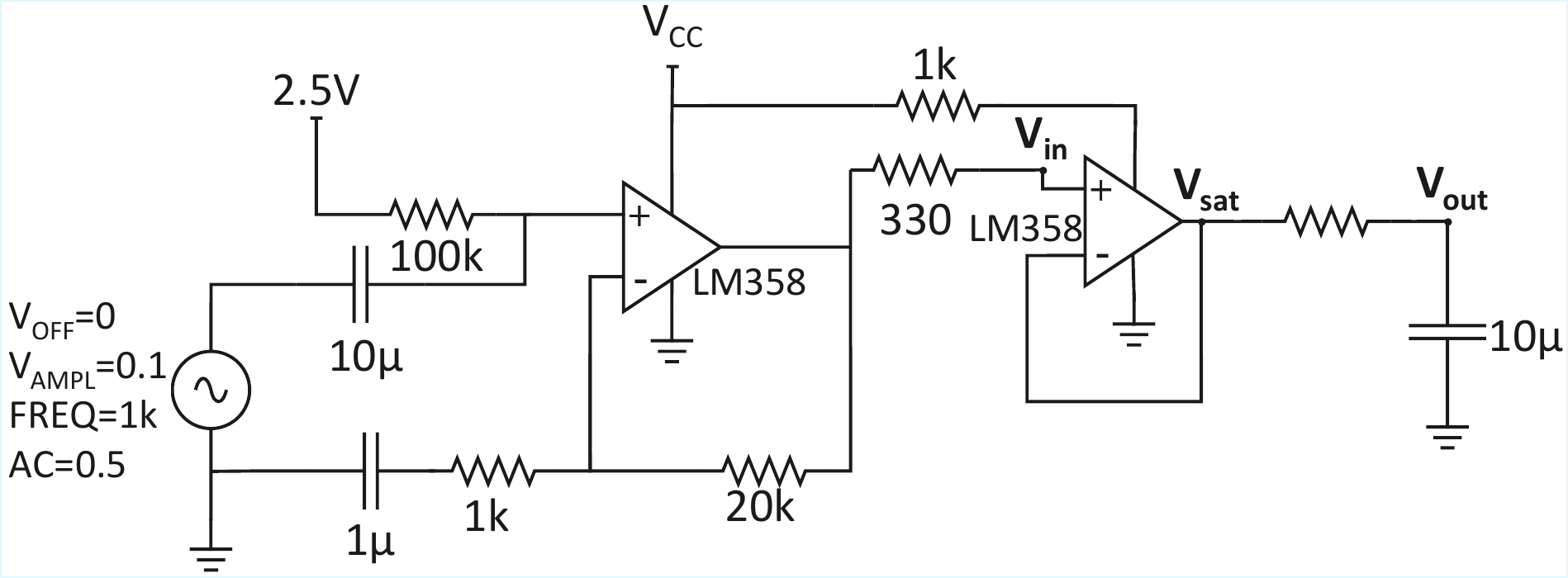}
    \caption{ The PSpice simulation circuit of a saturation Trojan in the amplifier. }
    \label{fig_amp1_simulate}
    \vspace{-5pt}
\end{figure}

We simulate the saturation amplifier circuit and introduce a $1k \Omega$ resistor between Vcc and the Vcc port of the amplifier as a Trojan. The specific circuit and results are shown in~\cref{fig_amp1_simulate}. When the amplitude of the injected signal exceeds $3V$, the output signal is cut off, inducing a DC offset.  When the amplitude of the input signal is lower than $3V$, the output is unaffected. The stealthiness is guaranteed because not all sensor operating ranges are full-scale. In our tests with the common operational amplifiers LM358 and LM318, the maximum output of the amplifier does not reach Vcc. Finally, the output signal of the low-pass filter, representing the induced offset signal, is about $0.5V$. For example, the range of an iron thermocouple is -210°C to 1200°C,  meaning a possible spoofing attack of  $ \mathrm{\pm 141^\circ C}$ can be introduced.

\subsubsection{The Envelope Detection Effect Simulation} \label{sec_sim_amp2}
\begin{figure}[H]
    \centering
    \includegraphics[width=1.0\linewidth]{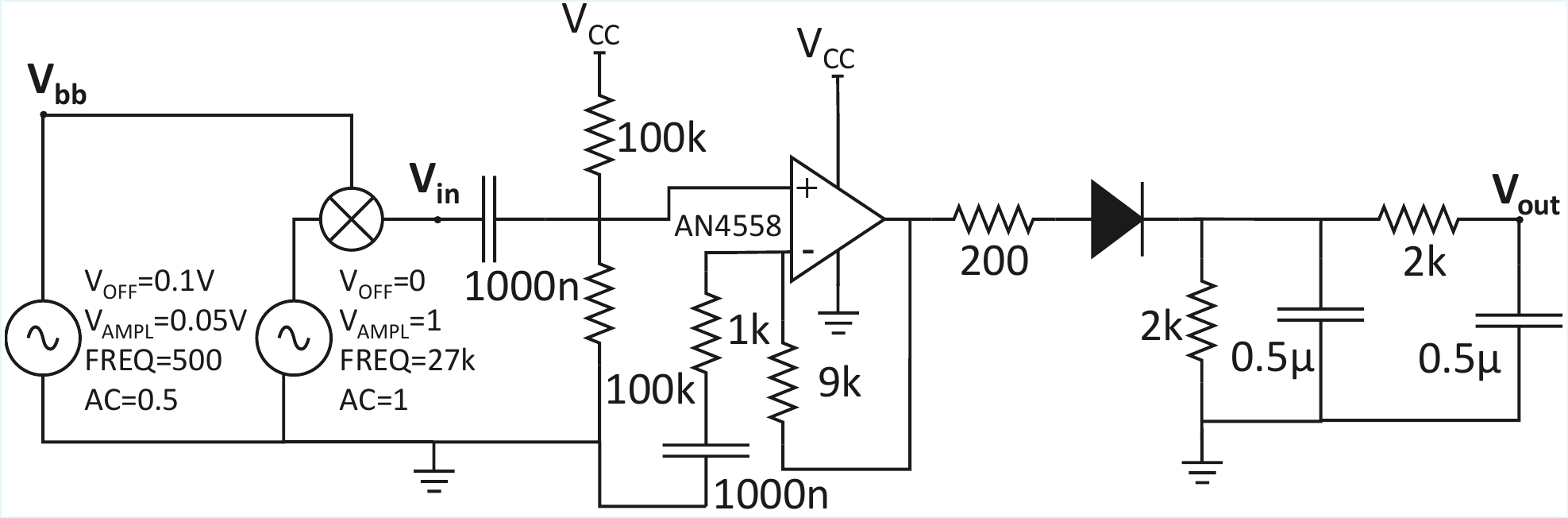}
    \caption{ The PSpice simulation circuit of an envelope detection Trojan in the amplifier. }
    \label{fig_amp2_simulate}
    \vspace{-5pt}
\end{figure}

\begin{figure}[H]
    \centering
    \includegraphics[width=1.0\linewidth]{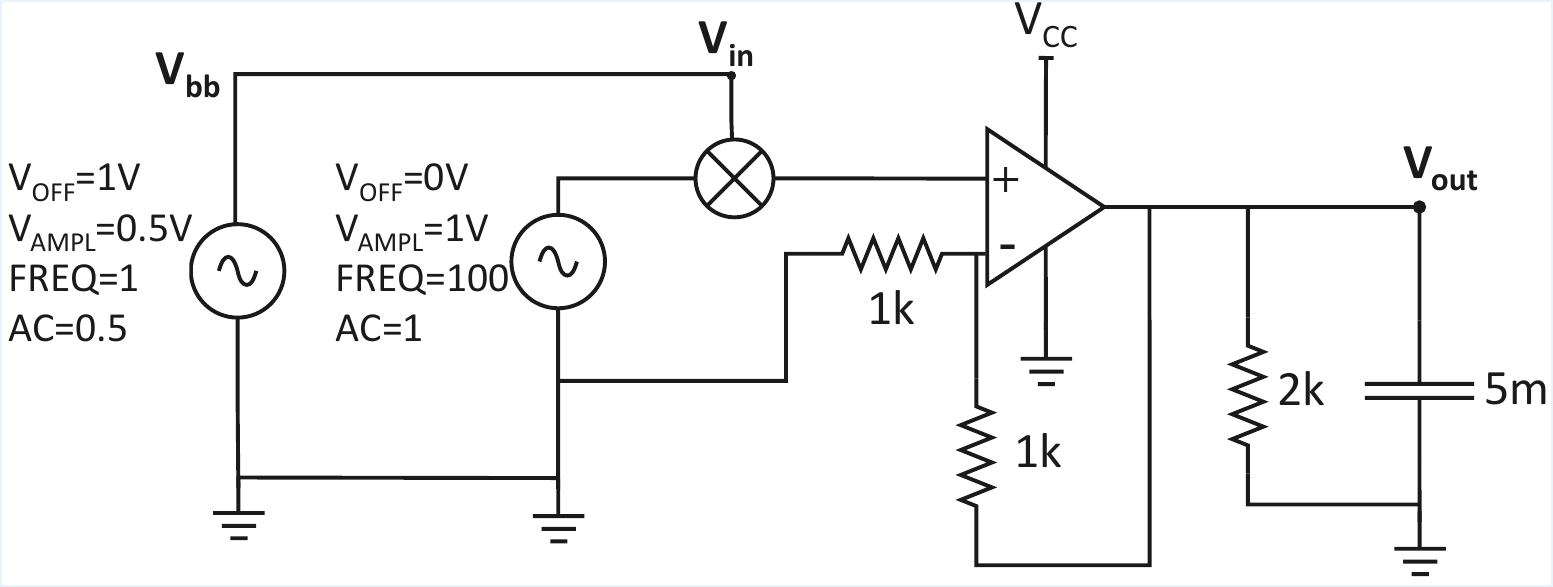}
    \caption{ The PSpice simulation circuit of an envelope detection Trojan in the amplifier without a diode. }
    \label{fig_amp2_simulate2}
    \vspace{-5pt}
\end{figure}

We simulate the envelope detection amplifier circuit and introduce a diode as the Trojan. The specific circuit is shown in~\cref{fig_amp2_simulate} and ~\cref{fig_amp2_simulate2}. 
In~\cref{fig_amp2_simulate}, the injected signal has a 27 kHz carrier modulated by a 500 Hz baseband signal. In~\cref{fig_amp2_simulate2}, the carrier and baseband frequencies are 100 Hz and 1 Hz, respectively.

The frequencies of these two signals determine the parameters of the subsequent demodulation circuit. The demodulation circuit recovers the baseband signal. For low-frequency signals, the capacitor, due to its high impedance at low frequencies, can be considered an open circuit and thus does not affect the circuit's performance.

\subsubsection{The Over-amplification Effect Simulation} \label{sec_sim_amp3}
\begin{figure}[H]
    \centering
    \includegraphics[width=0.8\linewidth]{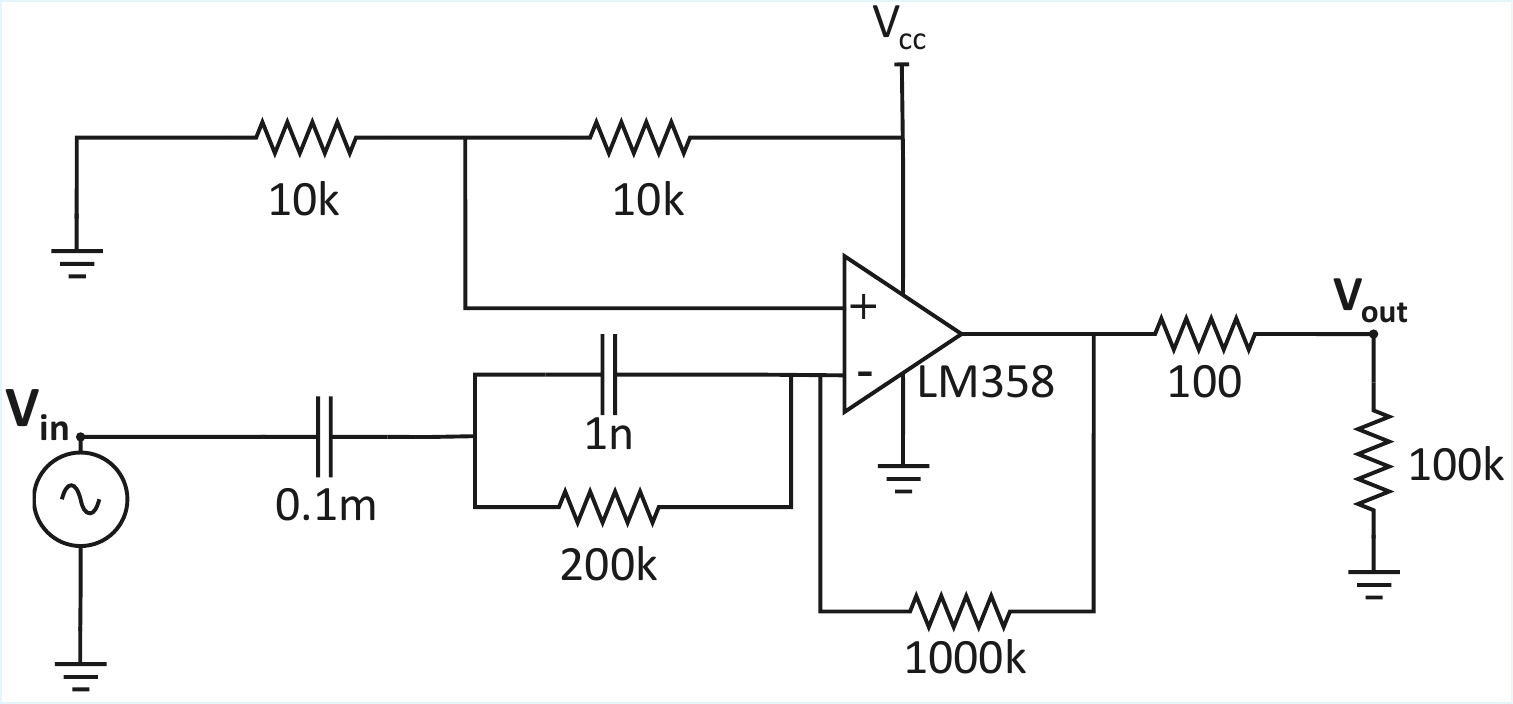}
    \caption{ The PSpice simulation circuit of an over-amplification Trojan in the amplifier. }
    \label{fig_amp3_simulate}
    \vspace{-5pt}
\end{figure}

Simulations depicted in~\cref{fig_amp3_simulate} utilize frequency sweeping to determine response curves. Results indicate amplification increases 200 times between 12 kHz and 13 kHz relative to low-frequency transmission, facilitating DoS attacks. Added capacitors function as open circuits for low-frequency inputs.  Unlike ideal models in prior expressions, practical bandwidth limitations suppress high-frequency signals, preventing unintended triggers.  Furthermore, in practical applications, the resistance introduced by wires or the addition of a small series resistor with the capacitor mitigates noise-induced false triggering. 

\subsection{The Trojan Test Details} \label{sec_app_detail}
\subsubsection{Light-triggered Trojan with Amplitude Mechanism}\label{sec_app_amp}
\begin{figure}[H]
 \centering
 \includegraphics[width=1.0\linewidth]{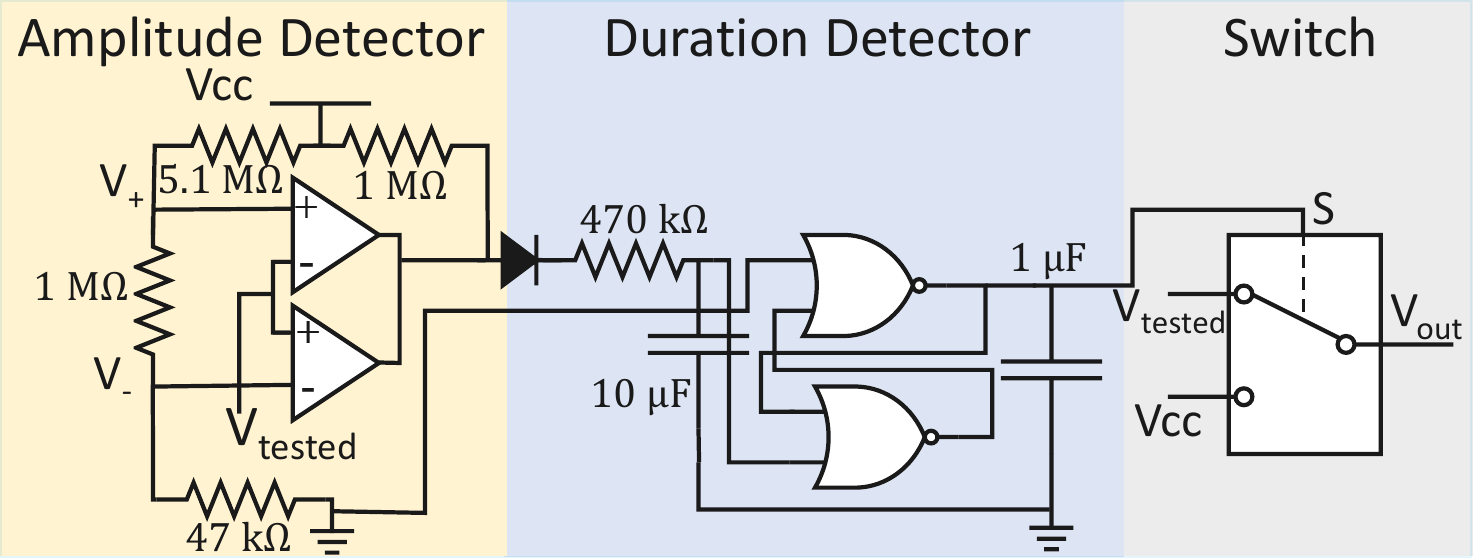}
 \caption{Schematics of an amplitude-based LRD Trojan in the GL5506.}
 \label{fig_case1_stru}
 \vspace{-5pt}
\end{figure}

Depicted in \cref{fig_case1_stru}, the Trojan circuit comprises a comparator, analog switch, latch, diode, capacitors, and resistors (3 SSI, 8 DC). A window comparator monitors real-time voltage (0.145 V to 0.970 V) mapped from light intensity. Sustaining signals within this range for 2.5 seconds charges the capacitor, toggling the analog switch to alter signal paths. Upon triggering, an electrical fault attack short-circuits critical wiring, potentially causing overheating and damage. Adapting for additional photoresistor transducers involves adding voltage dividers for detection and routing original sensor outputs to the switch, ensuring normal operation.


\subsubsection{Ultrasound-triggered Trojan with Frequency Mechanism}\label{sec_app_freq}
\begin{figure}[H]
 \centering
 \includegraphics[width=1.0\linewidth]{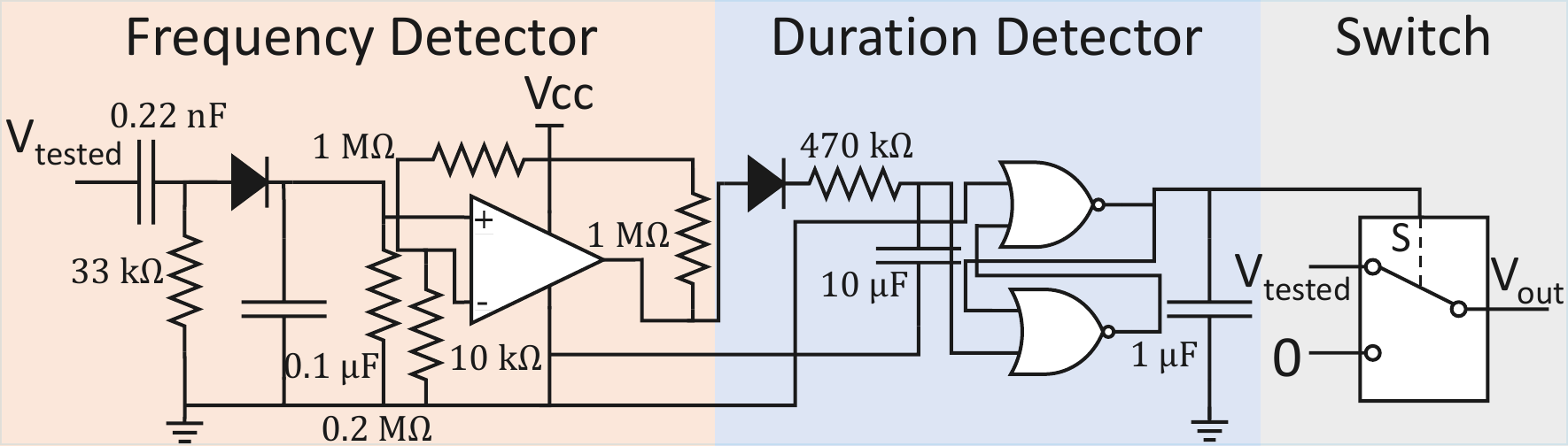}
 \caption{Schematics of a frequency-based LRD Trojan in the Sen0487.}
 \label{fig_case2_stru}
 \vspace{-5pt}
\end{figure}

Minimizing component count relative to band-pass alternatives, a high-pass filter ($33k\,\Omega$, $0.22\,\text{nF}$, $21,933\,\text{Hz}$) processes ultrasound signals, depicted in~\cref{fig_case2_stru}. Peak detectors extract amplitudes while parallel resistors prevent charge buildup. To mitigate 0.3 V leakage from filter imperfections, a comparator sets a 0.6 V threshold to prevent false triggers, augmented by duration detectors. The payload employs analog switches akin to~\cref{sec_light}.
In single transducer sensors lacking signal conditioning, low microphone outputs and diode drops impede peak extraction. Consequently, attackers insert amplifiers and feedback resistors to boost AC amplitudes, enabling Trojan functionality.
In addition, utilizing additional microphone transducers entails configuring basic ECM setups and routing analog switch $V_{test}$ inputs to original sensor outputs.




\subsection{Comparison with A2} \label{sec_app_A2}

SenTrojan differs from A2~\cite{yang2016a2} along four dimensions:

\begin{itemize}
    \item Trigger source: A2 is activated by internal digital states through a crafted program, whereas SenTrojan is triggered by external physical signals injected through the sensor. 

    \item Insertion point: A2 inserts malicious circuitry into unused layout areas, whereas SenTrojan targets sensor-associated analog front-end and PCB-level circuitry.

    \item Payload: A2 modifies a digital privilege flip-flop after activation. SenTrojan manipulates sensor outputs to induce PDoS, static spoofing, or dynamic spoofing.

    \item Validation: A2 uses fabricated malicious processors and software triggers. SenTrojan is evaluated on 14 sensors spanning six transduction principles, with system-level case studies.
\end{itemize}